\documentclass{article}
\usepackage{arxiv}

\usepackage{amsthm,amsfonts,amsmath,amssymb,empheq}
\usepackage{color}
\definecolor{Blue}{RGB}{0,0,255}
\usepackage[unicode, colorlinks, urlcolor=Blue, linkcolor=Blue, citecolor=Blue]{hyperref}
\usepackage{subcaption}

\usepackage[ruled,vlined]{algorithm2e}
\usepackage{textcomp}
\usepackage[normalem]{ulem}

\newtheorem{theorem}{Theorem}[section]
\newtheorem{proposition}{Proposition}[section]
\newtheorem{remark}[theorem]{Remark}
\theoremstyle{definition}

\begin{document}


\title{Numerical analysis of a self-similar turbulent flow in Bose--Einstein condensates}

\author{
 B.~V. ~Semisalov$^{1,*}$, V.~N.~Grebenev$^{2}$, S.~B.~Medvedev$^{2,1}$, S.~V.~Nazarenko$^{3}$\\
$^{1}$ Novosibirsk State University, Novosibirsk 630090, Russia,\\ 
$^{2}$ Federal Research Center for Information and Computational Technologies,\\ Novosibirsk
630090, Russia,\\
$^{3}$ Insitute de Physique de Nice, Universite  C$\hat o$te D'Azur,  Ave. Joseph Vallot, Nice 06100, France,\\
* Corresponding author: vibis@ngs.ru
}






\maketitle

\begin{abstract}
We study a self-similar solution of the  kinetic equation describing weak wave turbulence in  Bose--Einstein condensates. This solution presumably corresponds to an asymptotic behavior of a spectrum evolving from a broad class of initial data, and it features a non-equilibrium finite-time condensation of the wave spectrum $n(\omega)$ at the zero frequency $\omega$. The self-similar solution is of the second kind, and it satisfies boundary conditions corresponding to a nonzero constant spectrum (with all its derivative being zero) at $\omega=0$ and a power-law asymptotic
$n(\omega) \to \omega^{-x}$ at $\omega \to \infty \;\; x\in \mathbb{R}^+$.  Finding it amounts to solving a nonlinear eigenvalue problem, i.e. finding the value $x^*$ of the exponent $x$ for which these two boundary conditions can be satisfied simultaneously.
To solve this problem we develop a new high-precision algorithm based on Chebyshev approximations and double exponential formulas for evaluating the collision integral, as well as the iterative techniques for solving the integro-differential equation for the self-similar shape function.  This procedures allow to achieve a solution with accuracy $\approx 4.7 \%$ which is realized for $x^*  \approx 1.22$.
\end{abstract}




\section{Introduction}
An important role in understanding the turbulence in Bose--Einstein condensates is played by the statistical description of random nonlinear waves (wave turbulence) of a Bose gas \cite{dyachenko1992optical,nazarenko2011wave,semikoz1995kinetics,semikoz1997condensation,lacaze2001dynamical,connaughton2004kinetic}. The analysis of the Gross--Pitaevskii kinetic equation \begin{color}{black} with the cubic non-linearity in the right-hand side (RHS)\end{color} in the case of weak non-linearity and random phases leads to the following integro-differential \begin{color}{black}4-wave kinetic\end{color}  equation describing the evolution of the wave action spectrum $n_{\omega}(t)=n(\omega,t)$ (see e.g. \cite{dyachenko1992optical}),
\begin{eqnarray}\label{E1}
\frac{d}{d t}n_{\omega} &=& \omega^{-1/2}\int S(\omega,\omega_1,\omega_2,\omega_3)
n_{\omega}n_{1}n_{2}n_{3}\\
&&\left(n_{\omega}^{-1} + n_{1}^{-1} - n_{2}^{-1} - n_{3}^{-1}\right)\delta(\omega + \omega_1 - \omega_2
- \omega_3)d\omega_1d\omega_2d\omega_3. \nonumber
\end{eqnarray}
Here $\omega=k^2$ is the wave frequency, $k=|\mathbf{k}|$, $\mathbf{k}=(k_1,k_2,k_3)$ is the wave vector in Fourier space, $n_i=n(\omega_i,t)$,
$\omega_i=k_i^2$, $i=1,2,3$ and $\delta$ is the Dirac delta function. The {\it collision} integral in the \begin{color}{black} RHS \end{color} of (\ref{E1}) is taken over the positive values $\omega_{1}$, $\omega_{2}$, $\omega_{3}$. The
kernel of the integral reads
\begin{equation}
S(\omega,\omega_1,\omega_2,\omega_3) = \min\left(\sqrt{\omega},\sqrt{\omega_1},\sqrt{\omega_2},\sqrt{\omega_3}\right).
\end{equation}
Equation \eqref{E1} obeys the law of conservation of the total number of particles,
\begin{equation}
N =  2\pi \int_0^\infty  \omega^{1/2} \, n_{\omega} \, d \omega,
\end{equation}
and the total energy,
\begin{equation}
E = 2 \pi \int_0^\infty   \omega^{3/2}  \, n_{\omega} \, d \omega.
\end{equation}

Respectively, equation \eqref{E1} has two thermodynamic equilibrium solutions $n_{\omega} = C_1$ and $n_{\omega} =  C_2\omega^{-1}$ (where $C_1$ and $C_2$ are arbitrary positive constants) corresponding the the particle and the energy equipartition in the 3D $k$-space respectively. It also has two non-equilibrium stationary spectra of the power law form $n_{\omega}= C\omega^{-x}$; they are realized for \begin{color}{black}$x = 3/2$ and $x = 7/6$.\end{color}
These  are the Kolmogorov--Zakharov (KZ) spectra corresponding to constant fluxes of the  energy and the particles towards the high and low frequencies respectively \cite{dyachenko1992optical,nazarenko2011wave}.

The kinetic equation (\ref{E1}) was  studied numerically in
\cite{semikoz1995kinetics,semikoz1997condensation,lacaze2001dynamical,connaughton2004kinetic}  
to find evolving spectra arising from decay of an initial  data.
It was shown that the spectrum blows up to infinity (very high numerical values, to be precise) in finite time $t^*$ at $\omega=0$. In wave turbulence, such a transfer of particles toward the zero frequency is called the inverse particle cascade, and it is usually associated with the stationary inverse cascade KZ spectrum, which is a power-law solution with exponent $x=7/6$ \cite{dyachenko1992optical,nazarenko2011wave}. 
Numerical simulations of \cite{semikoz1995kinetics,semikoz1997condensation,lacaze2001dynamical,connaughton2004kinetic} did reveal the power-law behavior 
$n_{\omega}(t) \sim \omega^{-x^*}$ developing in the tail and invading the whole frequency range as
$t  \to t^*$. However, the observed exponents  $x^* \approx 1.24$  in \cite{semikoz1995kinetics,semikoz1997condensation} and $x^* \approx 1.2345$
in \cite{lacaze2001dynamical,connaughton2004kinetic} are both clearly different from the KZ exponent
$7/6$.
It was also noted that \begin{color}{black}there is a connection\end{color} between the observed behavior and the self-similar solutions. Such blow-up behavior is typical for the self-similarity of
the second kind. Indeed, by the Zeldovich--Raizer classification \cite{zel2002physics}, the self-similar solution is of the second kind if its similarity properties cannot be fully determined from a conservation law -- conservation of the total number of particles in our case. This is because the self-similar part of the evolving solution contains only a tiny part of the total number of particles. This is also reflected in the fact that the integral defining the respective invariant (particles) converges on the respective KZ spectrum at the limit toward which the cascade is directed -- limit $\omega \to 0$ in our case. Such a property is called the finite capacity of a spectrum. As a consequence, one cannot analytically determine the exponent $x^*$ of the asymptotic tail: one has to solve a ``nonlinear eigenvalue problem" numerically (see below).

Notice that the self-similarity  of the second kind is quite common in turbulence models, including the integro-differential wave kinetic equations \cite{galtier2000weak,connaughton2010dynamical,semikoz1995kinetics,semikoz1997condensation,lacaze2001dynamical,connaughton2004kinetic}, and local hydrodynamic and wave turbulence models represented by nonlinear partial differential equations (PDEs) \cite{connaughton2004warm,connaughton2003non,grebenev2013self,thalabard2015anomalous,galtier2019nonlinear}.  For the second-order PDE models, the problem of finding  $x^*$ boils down to finding a global heteroclinic bifurcation of the respective two-dimensional dynamical system \cite{grebenev2013self,thalabard2015anomalous}. Much less is known about the differential models of higher order, and there are practically no rigorous results about the self-similar solutions in the  integral kinetic equations (there is even no proof of their existence). In this case, one had  to resort to extracting information from the direct numerical simulations of the evolving kinetic equations, as it was done for the equation (\ref{E1})  in papers \cite{semikoz1995kinetics,semikoz1997condensation,lacaze2001dynamical,connaughton2004kinetic},  for a three-wave kinetic equation describing MHD wave turbulence in 
\cite{galtier2000weak} and for a simplified model three-wave kinetic equation in \cite{connaughton2010dynamical}. The only exception was a direct study  within the self-similar ansatz for the MHD wave turbulence undertaken in \cite{bell2017self}. The self-similar ansatz reduces the number of independent variables in the integro-differential equation to one, and its treatment requires setting up an iterative solution algorithm. This has proven to be a difficult task because the maps of such iterations are nonlinear and it is not possible to get a rigorous mathematical result about their contracting properties.

Past simulations  of the initial value problem governed by the integro-differential wave kinetic equations \cite{galtier2000weak,connaughton2010dynamical,semikoz1995kinetics,semikoz1997condensation,lacaze2001dynamical,connaughton2004kinetic} were very valuable as they pointed at the likely self-similar character of evolution and allowed to measure
the exponent $x^*$ of the asymptotic tail.  However, the simulations of the evolving spectrum are limited because the self-similar evolution of the second kind is expected to arise only very close to the blow-up time $t^*$ and only at frequencies which are very far from the initial ones. As a result, it is virtually impossible to show with any reasonable degree of accuracy that the evolving spectrum can be rescaled to collapse on a universal self-similar shape. Thus, the direct studies of within the self-similar ansatz are important for putting the self-similar formulation on a firm footing. 

The aim of the present paper is to find numerically the self-similar solution of equation~(\ref{E1}) directly using the self-similarity ansatz. \begin{color}{black} The key role in solving the associated eigenvalue problem is played by formulation of the correct boundary conditions\end{color}.
The results of solving the initial value problem indicate that, in addition to the power law asymptotic at large frequencies, the spectrum develops a plateau at small frequencies. Thus, our initial conjecture that the relevant self-similar solution must have (i) a power-law asymptotic at large $\omega$'s (with an exponent $x^*$ determined by the nonlinear eigenvalue problem)  and (ii) a ``shelf-like" structure at low $\omega$'s, implying that the spectrum at zero $\omega$ is finite and all its $\omega$-derivatives are zero. We will show that the {\it shelf-type} behavior of the spectrum in the vicinity of zero frequency is self-consistent by analyzing the contribution to the collision integral from the nonlocal interactions.  In addition, the physical meaning of the spectrum implies a requirement that the solution remains within the class of positive functions.

The present paper's approach is close in spirit to the study of Ref.~\cite{bell2017self} where a three-wave integro-differential equation (describing the MHD wave turbulence) was studied within the self-similar ansatz. However, the present paper is different in that it makes a special effort to increase accuracy of computing the collision integral by careful treating the vicinities of the singularities in the integrand kernel, as well as by explicitly measuring the error associated with the numerical self-similar solution. 
In particular, with an accuracy about $4.7\%$,  we find $x^*=1.22$ which is  quite close to the value $x^* = 1.24$ obtained by simulation of the initial value problem in Ref.~\cite{semikoz1995kinetics}
and
$x^* = 1.2345$
of Ref.~\cite{connaughton2004kinetic,lacaze2001dynamical}.  Considering the fact that the numerical errors were not reported in
the previous studies, we believe that the self-similar solution obtained in the present paper is the best result both in terms of the demonstration of existence of the solution to the stated nonlinear eigenvalue problem and the numerical accuracy of the obtained self-similar shape and the asymptotic exponent $x^*$.  

Development of a high-precision algorithm for computing the  collision integral presents a significant part of the present study. It is based on splitting the integration area into triangular, rectangular, trapezoidal and semi-infinite shapes, which are further mapped onto a square or a half-stripe where Chebyshev approximations and double exponential formulas are applied. The method's accuracy was benchmarked on typical spectrum shapes and it appears to be significantly higher that the ultimate accuracy of the self-similar solution. This is because the bottleneck for the solution accuracy was in the poor convergence properties of the iterations rather than in computing the collision integral. Thus, the developed method for computing the collision integral is likely to come useful in future  for tasks beyond finding the self-similar shapes, in particular for computing the non-stationary initial value problems.

\begin{color}{black}
Before we proceed it is worth noting that  the 4-wave kinetic equations have also been considered beyond the non-equilibrium Bose--Einstein condensation problem. For example, the well-known Hasselmann kinetic equation has been widely used to forecast the waves in oceans, including prediction of the fetched-limited (inhomogeneous in space) and anisotropic development of the wave fields. There are many studies (see, for example, \cite{korotkevich2008numerical,badulin2005self,badulin2017ocean,polnikov1997nonlinear,janssen2003nonlinear,resio1991numerical}) that treat this situation by considering the collision integral of higher dimension (including the angular dependencies) and by accounting for the dependence of the spectra on spatial coordinates, which requires introducing extra terms to the kinetic equation (describing motion of wavepackets with group velocities). Self-similar evolution in such setup was also studied (see, for example, the works of V. Polnikov and the group of V. Zakharov). However,  the considered self-similar solutions where of the first type rather than the second-type as in our paper.  

To solve numerically the kinetic equations that appear in operational forecasting, the Discrete Interaction Approximation (DIA), the Diffusion Approximation (DA), the Reduced Integral Approximation (RID) and some other methods have been used, \cite{van2006wrt,polnikov2002problem}. However, they give a rather rough approximation. Monte--Carlo methods work more accurately but very slowly. We can conclude that including extra effects, such as inisotropy and inhomogeneity, complicates the problem and makes it hard to discover nontrivial effects associated with the self-similarity of the second kind.

In our study we considered much more idealised situation of the wave fields which are isotropic and homogeneous in $k$-space. Due to it one can get rid of most integrals and pass to the 2-dimensional collision integral as in equation \eqref{E1}. For computing it we have developed very precise and fast methods. This allowed us to perform much more rigorous and accurate study.
\end{color}

\section{Statement of the problem}\label{S1}

We begin by  introducing  the self-similar ansatz for equation \eqref{E1}  by seeking the solution in the following form,
$$n_{\omega} = f(\eta) \tau^{-a},$$ where $\eta = \omega \tau^{-b}$, $\tau = t^* - t$. 

Denoting $x=\dfrac{a}{b}$, under condition $b =a-1/2> 0$ (meaning that the solution front moves toward small $\omega$'s),
equation \eqref{E1}  can be rewritten as an equation in only one (similarity) variable,
\begin{equation}
\label{C1_00} 
xf+\eta \frac{d
f}{d\eta}=\frac{1}{b}\biggl(A(f,\eta)+fB(f,\eta)\biggr),
\end{equation}
where
$$
A(f,\eta)=\eta^{-1/2}\int\limits_{\Delta_{\eta}} S\cdot
(f_2f_3f_c) ~d\eta_2 d\eta_3,~~~ B(f,
\eta)=\eta^{-1/2}\int\limits_{\Delta_{\eta}} S\cdot (f_2f_3 -
f_3f_c - f_2f_c) d\eta_2 d\eta_3,
$$
\begin{equation}
\label{C1_0}
S=\min\{\sqrt{\eta},\sqrt{\eta_2},\sqrt{\eta_2+\eta_3-\eta},\sqrt{\eta_3}\},
\end{equation}
$f_2=f(\eta_2)$, $f_3=f(\eta_3)$, $f_c=f(\eta_2+\eta_3-\eta)$,
$\Delta_{\eta}=\{(\eta_2,\eta_3):\eta_2>0,\eta_3>0,\eta_2+\eta_3>\eta\}$ is the domain of integration, see Fig.~\ref{fig1}.

To formulate the nonlinear eigenvalue problem, we have to complement the similarity equation ({\ref{C1_00}) with
boundary conditions at small and large values of $\eta$.

\subsection{Locality of interactions}
Appendix 1  studies convergence of the collision integral in the RHS of~(\ref{E1}) on the power spectrum
 $n_{\omega}=\omega^{-x}$ . The result is that this integral is convergent for $1<x<3/2$ and divergent otherwise.

Our initial proposition, to be justified {\it a posteriori}, is that the spectrum has two power-law asymptotics with exponent
$x^*$ in the locality range $(1,3/2)$ at large frequencies and with $x<1$ at small frequencies.
Note that these properties make the collision integral in (\ref{E1}) independent of the integration limits because, as shown in the appendix~1, the divergence for shallow spectra with $x<1$ occurs at the ultraviolet (infinite frequency) end of the integration range. This makes the self-similar ansatz self-consistent since the integral is convergent in this case on the full self-similar shape. Moreover, this setup allows one to establish the correct boundary condition at the left end of the $\eta$-range, as we will now show.

\subsection{Small--$\eta$ behavior}\label{NI}

From the previous numerical simulations of the initial value problem \cite{semikoz1995kinetics,semikoz1997condensation,lacaze2001dynamical,connaughton2004kinetic}, we get a hint that the solution must develop a ``shelf" at low frequencies, $n_\omega(t) \approx $~const for small $\omega$.
First of all, we need to define what we mean by small $\omega$. Let us suppose that the small $\omega$ asymptotic power law changes to the large $\omega$ asymptotic power law around some transitional frequency $\omega_{tr}(t)$.
As we will see below, the self-similar solution is defined up to a one-parametric family of scaling transformations, so we can always choose the transitional region so that $\eta_{tr}\sim1$. If so, according to our starting hypothesis
$f(\eta) \sim \eta^{-x_s}$ with $x_s <1$ for $\eta \ll 1$. Then, according to the results of the appendix~1, for  $\eta \ll 1$ the main contribution 
 to the integral in the RHS of \eqref{C1_00} is given by the integration over the
domain $\eta_{2,3} \sim 1 \gg \eta$, and equation \eqref{C1_00} can be rewritten as follows
\begin{color}{black}
\begin{equation}
\label{C10}
xf + \eta \frac{d
f}{d\eta} =  \frac{\tilde A}{b} +  \frac{\tilde B}{b}f,
\end{equation}
where 
\begin{equation}
\label{E10}
\tilde A = \int\limits_{\Delta_{0}} (f_{2}f_{3}f_{c})d\eta_2d\eta_3,~~~\tilde B = \int\limits_{\Delta_{0}} (f_{2}f_{3}-f_{3}f_{c}-f_{2}f_{c})d\eta_2d\eta_3.
\end{equation}
These integrals do not depend on $\eta$ and $f(\eta)$. Notice also that the domain of integration in \eqref{E10} can be taken equal to the whole first quadrant $\Delta_{0}$ in the plane $(\eta_2,\eta_3)$ due to the condition $\eta\ll 1$.
\end{color}

Equation~(\ref{C10}) can be easily integrated and the general solution reads
\begin{equation}\label{E10_1}
f(\eta) =  \frac{\tilde A}{b(x - \tilde B/b)} +  C\eta^{(\tilde B/b) - x},
\end{equation}
where $C$ is an arbitrary constant. \begin{color}{black}Taking into account the requirement $f(\eta)\geq0$ for the first integral expression in \eqref{E10}, one obtains $\tilde A>0$. This  
and $f(\eta) \ne \infty$\end{color} implies that  $x>\tilde B/b$ and 
$C=0$. Then for $\eta\ll 1$  we have an $\eta$-independent solution,
\begin{equation}\label{E11}
f(\eta) =  \frac{\tilde A}{(bx - \tilde B)}.
\end{equation}
In the other words,  $f(\eta)$ is a finite nonzero number and all of its derivatives are zero at $\eta=0$.
We shall refer to this property of $f(\eta)$ as a  shelf-type behavior.

Thus, we get $x_s =0$ which {\em a posteriori} justifies our initial hypothesis $x_s <1$. It should be emphasized that here one shouldn't confuse the powers $x$ and $x_s$: $x$ relates to the asymptotic $\eta\gg 1$ and $x_s$ relates to $\eta\ll 1$.

\subsection{Large--$\eta$ behavior}\label{Large}

Let us return to equation \eqref{C10}. Let us suppose  that the RHS of~\eqref{C10} tends to zero as $\eta\rightarrow\infty$
faster than each of the terms on the left-hand side (LHS), i.e. faster than $f(\eta)$. Then $xf + \eta df/d\eta \approx 0$ for $\eta \gg 1$ and $f(\eta)$ behaves as $\eta^{-x}$ as $\eta \to \infty$.
Substituting this to the RHS of \eqref{C10} and taking into account the collision integral convergence for
$1<x<3/2$, 
we see that the RHS behaves as $\sim \eta^{-3x +2}$, i.e. we confirm that it indeed decays faster than $f(\eta) \sim \eta^{-x}$ as $\eta \to \infty$. Thus, the asymptotic of the kind $\eta^{-x}$ at large $\eta$ is correct.

\subsection{Nonlinear eigenvalue problem}\label{BVP}
\label{S2}

Taking into account our findings about the small- and large-$\eta$ behaviors, we can now formulate the nonlinear eigenvalue problem. Let us consider the boundary value problem:
\begin{equation}
\label{C1} xf+\eta \frac{d
f}{d\eta}=\frac{1}{b}\biggl(A(f,\eta)+fB(f,\eta)\biggr),
\end{equation}
with boundary conditions
\begin{equation}
\label{C1_2} \frac{df}{d\eta}= 0\text{ at } \eta= 0 \quad
\hbox{and} \quad
xf+\eta \frac{d
f}{d\eta}\to 0\text{ for } \eta\to \infty.
\end{equation}
Our major hypothesis is that there exists a single value of the
 exponent $x=x^*$ for which the boundary value problem \eqref{C1}, \eqref{C1_2} is solvable.
 The nonlinear eigenvalue problem is formulated as finding $x^*$ (the latter being the eigenvalue in this case).

Notice that even though the eigenvalue $x^*$ is unique, the solution $f(\eta)$ of the problem \eqref{C1}, \eqref{C1_2} admits the  group of scaling transformations
\begin{equation}
\label{C1_3} \eta\rightarrow\tilde{\eta}=C\eta,~~~f\rightarrow\tilde{f}=f/C.
\end{equation}

\section{Numerical setup}

For  adapting the problem to numerical simulations, we introduce  an interval of positive values $[\eta_{min}, \eta_{max}] \subset (0,\infty)$, set $f(\eta) = C_0$ for $\eta<\eta_{min}$ and $f(\eta) = C_{\infty}\eta^{-x}$ for $\eta>\eta_{max}$ where
$\eta_{min} \ll \eta_{max}$. Here $C_0$, $C_{\infty}$ are positive constants. We will refer to $[\eta_{min},\eta_{max}]$ as the intermediate region and instead of studying the problem on $[0,\infty)$, we consider equation \eqref{C1} defined on $[\eta_{min},\eta_{max}]$ supplemented with the boundary conditions
\begin{equation}
\label{C2}
\frac{df}{d\eta}\bigg|_{\eta=\eta_{min}}=0,~~~xf+\eta \frac{d
f}{d\eta}\bigg|_{\eta=\eta_{max}}=0.
\end{equation}
The boundary value problem \eqref{C1}, \eqref{C2} is again a nonlinear eigenvalue problem.
Since the original problem admits the  group of scaling transformations,
we can fix the right boundary $\eta_{max}$ and look for the left boundary $\eta_{min}$ providing solvability of the eigenvalue problem for equation \eqref{C1} with boundary conditions \eqref{C2}. For numerical simulations we will take $\eta_{max}\in[10,100]$ and with suitable scaling we can search $\eta_{min}$ on the interval $(0,\eta_{max}/10]$.
A solution of the original problem \eqref {C1}, \eqref {C1_2} is obtained from \eqref{C1}, \eqref{C2} by the trivial continuation
\begin{equation}
\label{C2_1}
f(\eta) = f(\eta_{min})\text{ for }\, 0\leq \eta\leq\eta_{min},\quad
f(\eta)=C_{\infty}\eta^{-x}\text{ for }\, \eta_{max}\leq \eta<\infty,
\end{equation}
where the constant $C_{\infty}$ follows from the continuity at $\eta_{max}$,
\begin{equation}
\label{Cinf}
C_{\infty}=f(\eta_{max})\eta_{max}^x~\text{or}~C_{\infty}=\frac{-A(f,\eta_{max})\eta_{max}^x}{B(f,\eta_{max})}.
\end{equation}
Here we used the second boundary condition \eqref{C2} and equation \eqref{C1}.
The formulas \eqref{C2_1} will be used in calculating the integrals $A(f,\eta)$, $B(f,\eta)$
in \eqref{C1}. Specifically, an iterative procedure will be applied for finding a solution of  \eqref{C1},
\eqref{C2} such that $f(\eta)$, determined on the previous iteration for $\eta\in[\eta_{min},\eta_{max}]$, is used to calculate
$C_{\infty}$ thereby extending $f(\eta)$ onto the interval $[0,\infty)$ and then substituting the resulting $f(\eta)$ into the RHS of equation \eqref{C1} to calculate  $A(f,\eta)$, $B(f,\eta)$ of the current iteration.

\begin{figure}[ht]
\centering
\includegraphics[scale=0.37]{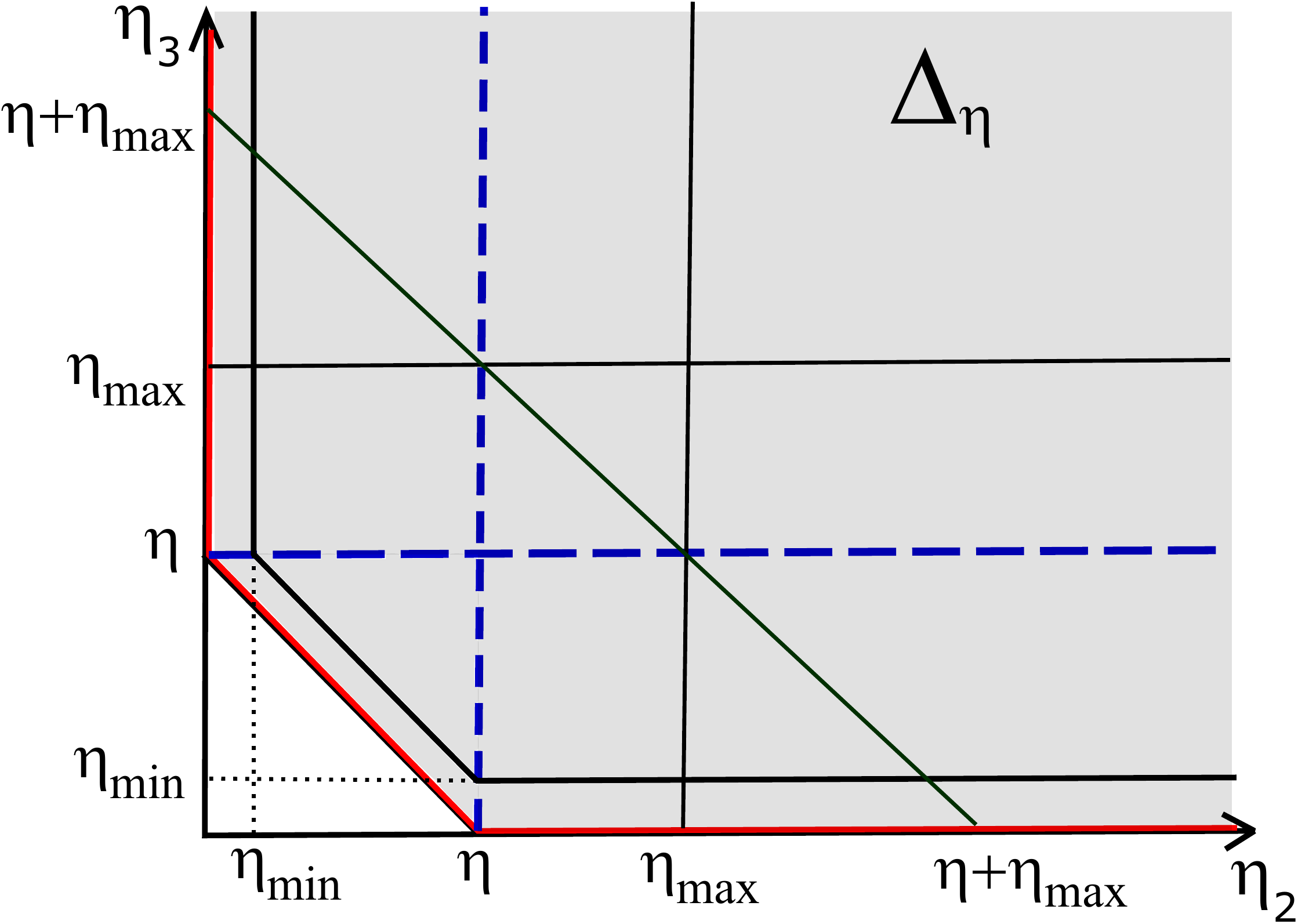}
\caption{The domain of integration $\Delta_{\eta}$ (shadowed). Solid lines show the borders $\eta_{min}$, $\eta_{max}$ for $\eta_2$, $\eta_3$, $\eta_2+\eta_3-\eta$. Dashed lines show the discontinuity of integrand's derivative due to presence of function ``$\min$''. Red
lines along the boundary show the location of possible singularity of integrand near zero values of $\eta_2$, $\eta_3$ and $\eta_2+\eta_3-\eta$ }\label{fig1}
\end{figure}

To solve the problem \eqref{C1}, \eqref{C2} numerically, we will decompose the integration domain $\Delta_{\eta}$ taking into account the 
singularities of the integrand. In particular, the matching conditions of the form \eqref{C2_1} at the points $\eta_{min}$,
$\eta_{max}$ in combination with \eqref{C2} ensure that $f(\eta)\in C^1(\mathbb{R}^+)$. However, the second and higher
order derivatives at these points can be discontinuous functions. The kernel $S=S(\eta_2,\eta_3)$ of the integral operators $A(f,\eta)$, $B(f,\eta)$ is a continuous function but the derivatives of $S(\eta_2,\eta_3)$ are discontinuous functions on the lines $\eta_2=\eta$, $\eta_3=\eta$ of the plain $(\eta_2,\eta_3)$, i.e. $S(\eta_2,\eta_3)\notin C^1(\Delta_{\eta})$. For fast calculations of
$A(f,\eta)$, $B(f,\eta)$ with an exponentially small  errors, we will use a decomposition of $\Delta_{\eta}$ along the
lines of discontinuity of the integrand derivatives so that inside of each subdomain $\Omega$ of the decomposition up to its border these functions belong to the class $C^{\infty}(\Omega)$. We decompose $\Delta_{\eta}$ into triangles, rectangles and trapeziums along the following lines (see
Fig.~\ref{fig1}):

$\eta_2=\eta$, $\eta_3=\eta$ (where the derivatives of the kernel $S(\eta_2,\eta_3)$ are broken);

$\eta_2=\eta_{min}$, $\eta_3=\eta_{min}$, $\eta_2+\eta_3=\eta_{min}$ (where the  $\eta_{min}$-matching  is done for  functions
$f_2$, $f_3$, $f_{c}$);

$\eta_2=\eta_{max}$, $\eta_3=\eta_{max}$,
$\eta_2+\eta_3=\eta_{max}$ (where the $\eta_{max}$-matching is done for functions $f_2$, $f_3$, $f_{c}$).

\begin{remark}
\label{rem2}
Formulas \eqref{C2_1} give a suitable realization of solutions of the original problem such that  $f(\eta)$ is 
constant near $\eta=0$. However, the intermediate iterations to the solution could be singular functions  for  $\eta \ll 1$.
Indeed, the conducted experiments  show that the numerical solutions  admit an oscillatory/peaked behavior for $\eta_{min} \ll 1$.
Nevertheless, we demonstrate that these oscillations can be damped by finding  the right eigenvalue $x^*$--roughly a value that would correspond 
to the zero-value constant $C$ in  \eqref{E10_1}. The red line presented in Fig.~\ref{fig1} indicates where such singularities of
numerical solutions can appear. Along this line the kernel $S$ also has the square-root singularity.
\end{remark}

\section{Calculation of  integrals $A(f,\eta)$ and $B(f,\eta)$}
\label{intComp}

For higher-order  (exponential) convergence of the cubature formulas applied for computing the integrals of the RHS of
equation \eqref{C1},  spectral approximations of the integrands are used. We will employ the
Chebyshev polynomials which will enable us to write  explicit formulas for the nodes of cubatures and explore the Fast Fourier Transform (FFT) to get the coefficients of the integrand expansion as polynomial series.

Throughout this section, it is assumed that $\eta \in [\eta_{min},\eta_{max}]$. We denote by $g(\eta_2,\eta_3)$ the integrand
of the operator $A(f,\eta)$. Notice that the formulas obtained below for the integration of $g(\eta_2,\eta_3)$ can be immediately applied for computing the integral $B(f,\eta)$.

Let $\Omega$ be a bounded subdomain of the decomposition of $\Delta_{\eta}$. To integrate  function $g(\eta_2,\eta_3)$ over $\Omega$,
we introduce a reference square domain $R_{sq}=\{(y,z):-1\leq y,z\leq 1\}=[-1,1]^2$. It is always possible to change the variables
$(\eta_2,\eta_3)$ to $(y,z)$ due to  existence of a $C^{\infty}$--mapping $\mathfrak{F}:\Omega\rightarrow R_{sq}$. Let
$g_{\Omega}(y,z)=g(\mathfrak{F}(\eta_{2},\eta_{3}))$, where $(\eta_{2},\eta_{3})\in\Omega$. We introduce the grid with nodes $(y_k,
z_m)\in R_{sq}$, where $y_k=\cos\frac{(2k-1)\pi}{2K}$, $z_m=\cos\frac{(2m-1)\pi}{2M}$ are zeroes of the Chebyshev polynomials
$T_K(y)=\cos(K\arccos y)$, $T_M(z)=\cos(M\arccos z)$, $k=1,...,K$, $m=1,...,M$. The decomposition of $\Delta_{\eta}$ together with the
grid $(\eta_2^k,\eta_3^m)=\mathfrak{F}^{-1}(x_k,y_m)$ for each subdomain $\Omega$ are shown on  Fig.~\ref{fig2}. For the unbounded subdomains,
the reference domains, the grid points and the mapping $\mathfrak{F}$ are described below.

\begin{figure}[ht]
\centering
\includegraphics[scale=0.32]{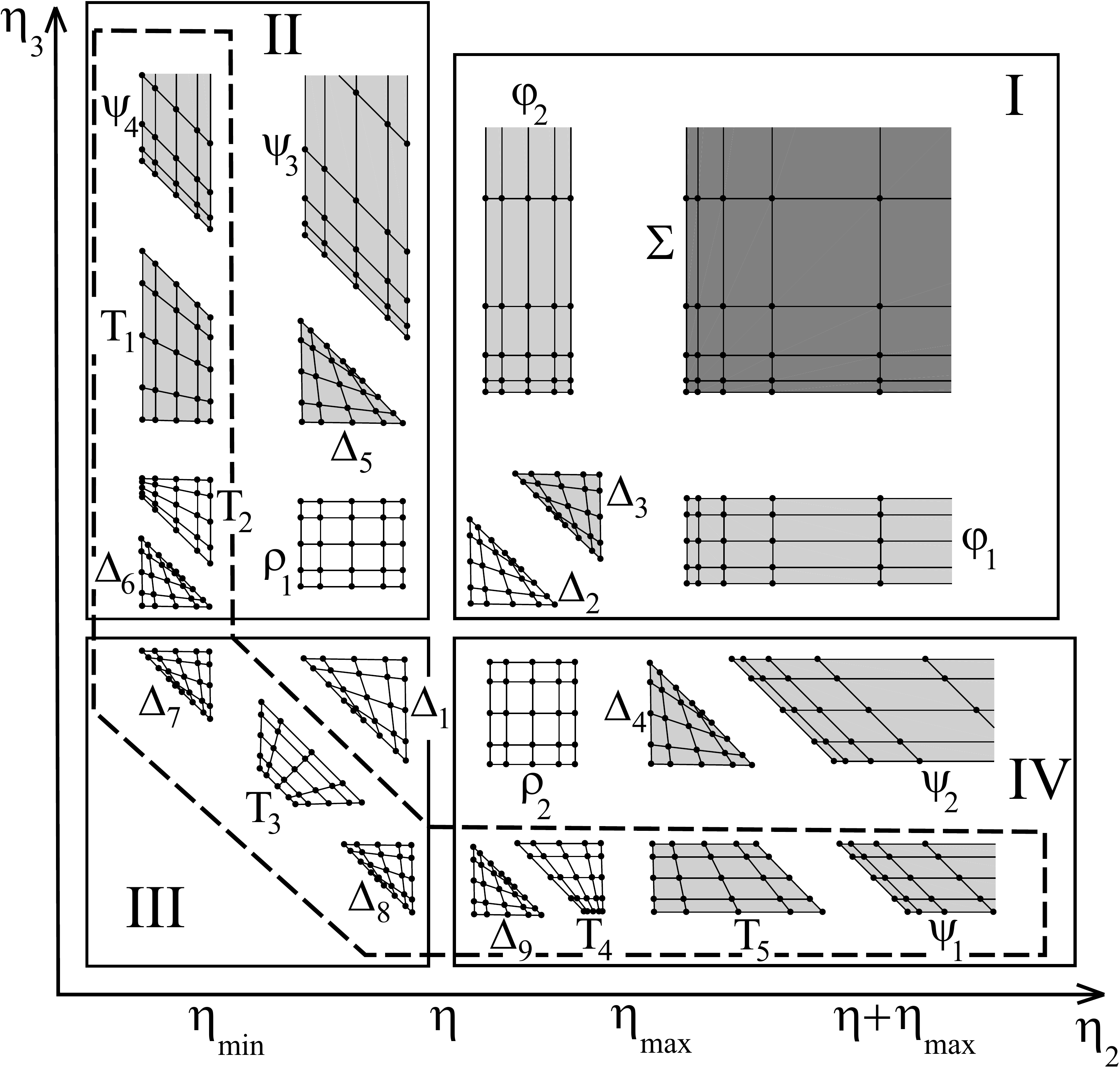}
\caption{Decomposition of the domain $\Delta_{\eta}$ used for integration. Triangular ($\Delta_k$), trapezoid (T$_k$), rectangular
($\rho_k$), infinite trapezoid ($\psi_k$) and infinite right-angular ($\varphi_k$, $\Sigma$) subdomains are shown. The nodes of
integration of $5\times 5$ meshes are marked with dots. ``I --IV'' are subdomains where  the kernel $S$ equals to the first, second,
third and fourth component of the minimum in \eqref{C1_0} correspondingly. White
subdomains are the ones were the integrand includes the values of $f$ with $\eta<\eta_{max}$ only; in light-gray subdomains the integrand
uses partially the asymptotic $\sim\eta^{-x}$  with $\eta\geq\eta_{max}$ and partially values of $f$ with $\eta<\eta_{max}$; dark-gray
subdomain $\Sigma$ uses the asymptotic $\sim\eta^{-x}$ only. Integrals over subdomains inside dashed region use the continuation of
$f(\eta)$ from $[\eta_{min},\eta_{max}]$ to $[0,\eta_{min}]$ by the constant value $f(\eta_{min})$}\label{fig2}
\end{figure}

To continue, we notice that the integrand $g_{\Omega}(y,z)$ is infinitely smooth inside of the subdomains $\Omega$. However, for subdomains located near the boundary of $\Delta_{\eta}$ (see red lines in Fig.~\ref{fig1}) $g_{\Omega}(y,z)$ has singularities near the  boundary of $\Omega$, see Appendix~3.

Let us first describe the approach for computing $A(f,\eta)$ and $B(f,\eta)$ when $g_{\Omega}(y,z)\in C^{\infty}(\overline{\Omega})$ is an analytic function over both of its variables. For $g_{\Omega}(y,z)$ we use the following approximation,
\begin{equation}
\label{C3}
g_{\Omega}(y_k,z)\approx\sum\limits_{m=0}^M a_{mk}
T_m(z),~~~a_{mk}=\frac{c_m}{\pi}\int\limits_{-1}^1\frac{g_{\Omega}(y_k,z)T_m(z)}{\sqrt{1-z^2}}dz,
\end{equation}
with $c_0=1,~c_m=2~\forall m>0$ and exponential rate of the convergence, see chapters 7,~8 in \cite{trefethen2019approximation}. Notice that the coefficients $a_{km}$ can be obtained from
$g_{\Omega}(y_k, z_m)$ by using the FFT. Let $I_g(y)=\int\limits_{-1}^1g_{\Omega}(y,z)dz$. With this, we have 
\begin{equation}
\label{C3a}
I_g(y_k)=\int\limits_{-1}^1 g_{\Omega}(y_k,z)dz\approx I_g^{M}(y_k)= \sum\limits_{m=2}^M a_{mk} \mathcal{T}_m,~~~
\mathcal{T}_m=\int\limits_{-1}^1T_m(z)dz,
\end{equation}
where we use that $\mathcal{T}_0=\mathcal{T}_1=0$,
$$
\mathcal{T}_k =\frac{1+\cos k\pi}{1-k^2},~k=2,3,...K.
$$
Notice also that $\mathcal{T}_m=0$ for all odd $m$. To integrate $g_{\Omega}(y,z)$ over
$R_{sq}$, we use the following approximation,
$$\int\limits_{R_{sq}}g_{\Omega}(y,z)ds_R=\int\limits_{-1}^1\biggl(\int\limits_{-1}^1g_{\Omega}(y,z)dz\biggr) dy=\int\limits_{-1}^1I_g(y)
dy\approx
\sum\limits_{k=2}^K A_k\mathcal{T}_k,$$
where $ds_R$ is the element of area of $R_{sq}$.  Here, $A_k$ denotes the coefficients of Chebyshev expansion of the function $I_g(y)$. These
coefficients have the form similar to $a_{mk}$ in \eqref{C3} and they can be determined by applying the FFT to the set of values
of $I_g(y_k)$, $k=1,...,K$ which are given by \eqref{C3a}. 

This approach goes back to Clenshaw, Curtis and Gentleman, see \cite{clenshaw1960method,gentleman1972implementing} for  details. The error estimates for
the integration are discussed in \cite{weideman2007kink}. Below, we modify this approach for the case of integrals with singularities described in
Remark~\ref{rem2}. To use the above-mentioned formulas for integration of the function $g(\eta_2, \eta_3)$ over a subdomain $\Omega$, the
function $g_{\Omega}(y,z)$ should be replaced by $g_{\Omega}(y,z)J_{\Omega}(y,z)$ where $J_{\Omega}(y,z)$ denotes the Jacobian of the
inverse mapping $\mathfrak{F}^{-1}$.
While applying the described method for computing the integrals the following features appear:
(1) non-regular shape of the subdomains; (2) the presence of unbounded subdomains; (3) the appearance of the singularity $\eta^{-p}$, $p > 0$ of the integrand on the lines $\eta_2=0$, $\eta_3=0$, $\eta_2+\eta_3-\eta=0$.  We will now discuss these features in more details.

\subsection{Feature 1: non-regular shape of the subdomains}\label{F1}

In order to perform the integration over subdomains having non-regular shapes, we have to specify $C^{\infty}$-mappings $\mathfrak{F}$.
For triangular subdomains the mapping is $\mathfrak{F}=\psi\circ\varphi$, where $\varphi$ maps a given triangle from
$(\eta_2,\eta_3)$-plain to the reference triangle $R_{tr}$ and $\psi$ maps $R_{tr}$ to $R_{sq}$, see Fig.~\ref{fig2_2} and
\cite{hossain2014generalized} for details.
\begin{figure}[hbt]
\centering
\includegraphics[scale=0.43]{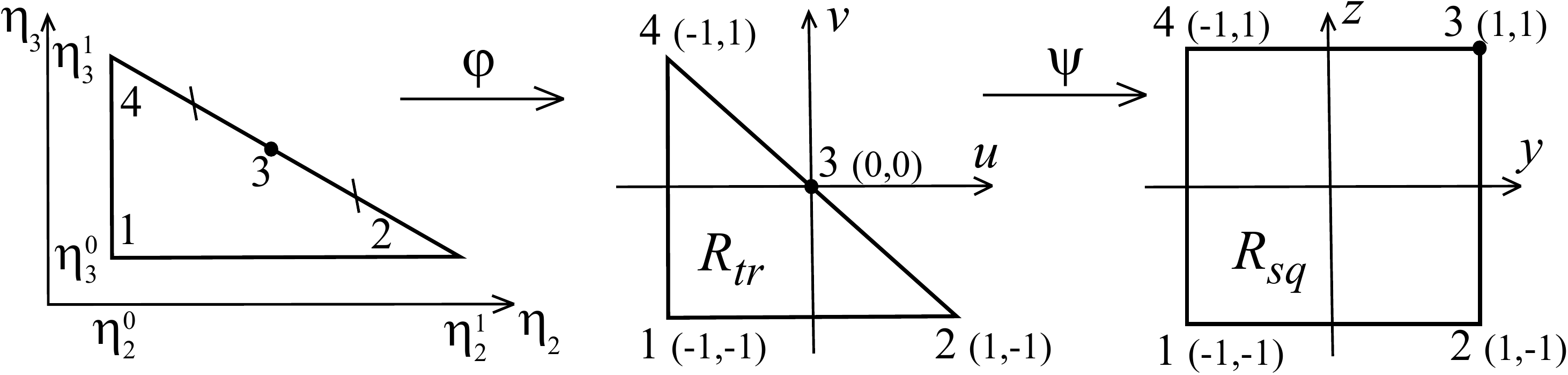}
\caption{Mapping of the triangular subdomains of $\Delta_{\eta}$ to reference square $R_{sq}$,
$\mathfrak{F}=\psi\circ\varphi$}\label{fig2_2}
\end{figure}

The general formulas for the mapping $\mathfrak{F}^{-1}$ and its Jacobian in the case of triangular subdomains are
\begin{equation}\label{C5}
\eta_2=\frac{\alpha}{4}(3y-z(y+1)-1)+\beta,~~~\eta_3=\frac{\gamma}{4}(3z-y(z+1)-1)+\theta,~~~J_{\Omega}(y,z)=\frac{\alpha\gamma}{4}(2-y-z),
\end{equation}
where (see Fig.~\ref{fig2_2}) $\alpha=\dfrac{\eta_2^1-\eta_2^0}{2}$, $\beta=\dfrac{\eta_2^1+\eta_2^0}{2}$,
$\gamma=\dfrac{\eta_3^1-\eta_3^0}{2}$, $\theta=\dfrac{\eta_3^1+\eta_3^0}{2}$. In our case $\alpha=\gamma$. For the subdomains
$\Delta_1,\dots,\Delta_9$ (see Fig.~\ref{fig2}) the parameters $\alpha$, $\beta$, $\gamma$ and $\theta$ are given in Table~\ref{tabTriangPar}.
\begin{table}[!hbt]
\caption{Values of the parameters $\alpha,\dots,\theta$ for $\Delta_1,\dots,\Delta_9$. The signs $\pm$ and $\mp$ correspond to the order of
subscripts in the notations $\Delta_{2,3}$, $\Delta_{6,7}$, $\Delta_{8,9}$}
\label{tabTriangPar}
\begin{center}
   \begin{tabular}{c|c|c|c|c|c|c}  \hline
Subdomain   & $\Delta_{1}$ & $\Delta_{2,3}$ & $\Delta_4$ & $\Delta_5$ & $\Delta_{6,7}$  & $\Delta_{8,9}$ \\
\hline & & & & & & \\
$\alpha=\gamma$ & $\frac{\eta_{min}-\eta}{2}$ & $\pm\frac{\eta_{max}-\eta}{2}$ & $\frac{\eta-\eta_{min}}{2}$ &
$\frac{\eta-\eta_{min}}{2}$ & $\pm\frac{\eta_{min}}{2}$ &  $\mp\frac{\eta_{min}}{2}$ \\
$\beta$  & $\frac{\eta_{min}+\eta}{2}$ & $\frac{\eta_{max}+\eta}{2}$ & $\frac{2\eta_{max}+\eta-\eta_{min}}{2}$ &
$\frac{\eta+\eta_{min}}{2}$ & $\frac{\eta_{min}}{2}$ & $\frac{2\eta\mp\eta_{min}}{2}$ \\
$\theta$ & $\frac{\eta_{min}+\eta}{2}$ & $\frac{\eta_{max}+\eta}{2}$ & $\frac{\eta+\eta_{min}}{2}$   &
$\frac{2\eta_{max}+\eta-\eta_{min}}{2}$ & $\frac{2\eta\pm\eta_{min}}{2}$ &  $\frac{\eta_{min}}{2}$  \\[1ex]
\hline
\end{tabular}
\end{center}
\end{table}

The explicit form of  $\mathfrak{F}^{-1}$ for the trapezoid subdomains $T_1,\dots,T_5$ together with the Jacobians $J_{T1},\dots,J_{T5}$ are
given in Appendix~2.
The rectangles $\rho_{1,2}$ can be mapped on $R_{sq}$ by using a linear change of variables. The Jacobians of the inverse mappings are
$J_{\rho1}=J_{\rho2}=\dfrac{(\eta-\eta_{min})(\eta_{max}-\eta)}{4}$. It is important to notice that all the Jacobians of the constructed mappings
 are $C^{\infty}$--smooth functions.

\subsection{Feature 2: unbounded  subdomains}\label{F2}

In order to integrate $g(\eta_2,\eta_3)$ over the unbounded subdomains $\psi_1$--$\psi_4$, $\varphi_1$, $\varphi_2$ and $\Sigma$, we use the
formulas (3.4)--(3.6) from \cite{takahasi1974double}.
Firstly we map the subdomains $\psi_{1,2}$, $\varphi_1$ into the reference stripe $\tilde{R}_{st}=\{(\tilde{y},\tilde{z}):
\tilde{y}\in[0,\infty), \tilde{z}\in[-1,1]\}$ and the subdomains $\psi_{3,4}$, $\varphi_2$ into another reference stripe
$\widehat{R}_{st}=\{(\widehat{y},\widehat{z}): \widehat{y}\in[-1,1], \widehat{z}\in[0,\infty)\}$. For these mappings we use again the
notation $\mathfrak{F}$. The formulas for $\mathfrak{F}^{-1}$ and the Jacobians $J_{\psi 1}$--$J_{\psi 4}$, $J_{\varphi 1}$,
$J_{\varphi 2}$ (which are again $C^{\infty}$--smooth functions) are given in Appendix~2.

We give the formulas for integration over the strip $\psi_3$ as an example. Let $g_{\psi3}=g\circ
\mathfrak{F}$ and the integral of $g(\eta_2,\eta_3)$ over $\psi_3$ can be written as
$$
I_{\psi3}=\int\limits_{\psi
3}g(\eta_2,\eta_3)d\eta_2d\eta_3=\int\limits_{-1}^1\biggl(\int\limits_{0}^{\infty}g_{\psi3}(\tilde{y},\tilde{z})J_{\psi3}(\tilde{z})d\tilde{z}\biggr)
d\tilde{y}=\int\limits_{-1}^1G(\tilde{y}) d\tilde{y}.
$$
Let $\tilde{y}$ be equal to the Chebyshev node $\tilde{y}_k=\cos\frac{(2k-1)\pi}{2K}$, $k\in\{1,...,K\}$. We use the change of variable
$\tilde{z}(u)=\exp\bigl(\dfrac{\pi}{2}\sinh u\bigr)$ with the Jacobian
$J_T(u)=\dfrac{\pi}{2}\tilde{z}(u)\cosh(u)$ where $u\in\mathbb{R}$. Then the integral reads
\begin{equation}
G(\tilde{y}_k)=\int\limits_{0}^{\infty}g_{\psi3}(\tilde{y}_k,\tilde{z})J_{\psi3}(\tilde{z})d\tilde{z}=\int\limits_{-\infty}^{\infty}
g_{\psi3}(\tilde{y}_k,\tilde{z}(u))J_{\psi3}(\tilde{z}(u))J_T(u)du
\end{equation}
and then it can be approximated by the Double Exponential Formula (see \cite{takahasi1974double})
\begin{equation}\label{Tak1}
G(\tilde{y}_k)\approx\mathcal{G}(\tilde{y}_k,h,N_t)=\dfrac{\pi}{2}h\sum\limits_{n=-N_t}^{N_t}g_{\psi3}(\tilde{y}_k,\tilde{z}(nh))J_{\psi3}\bigl(\tilde{z}(nh)\bigr)(\cosh(nh))\tilde{z}(nh),
\end{equation}
where $\tilde{z}(nh)=\exp\bigl(\dfrac{\pi}{2}\sinh nh\bigr)$,  $h$ is the step of integration, $nh$ denotes the nodes of the mesh.
Using the results by Takahasi and Mori (see~\cite{takahasi1974double}), we can formulate
\begin{proposition}
Let $g_{\psi3}(\tilde{y}_k,\tilde{z})$ be analytic function with respect to the variable $\tilde{z}$ then $G(\tilde{y}_k)\approx
\mathcal{G}(\tilde{y}_k,h,\infty)\approx \mathcal{G}(\tilde{y}_k,h,N_t)$. If $$h\approx \dfrac{1}{N_t}\log(4\theta N_t),$$ then
for $N_t\rightarrow\infty$
\begin{equation}\label{C7}
|G(\tilde{y}_k)- \mathcal{G}(\tilde{y}_k,h,\infty)|\sim|\mathcal{G}(\tilde{y}_k,h,\infty)-
\mathcal{G}(\tilde{y}_k,h,N_t)|\sim\exp\left\{-\frac{2\pi\theta}{\log(4\theta N_t)}N_t\right\},
\end{equation}
where $\theta>0$ is the distance from the real axis to the nearest singularity
in $\mathbb{C}$
of the function $g_{\psi3}(\tilde{y}_k,\tilde{z})$.
\end{proposition}
The proof is based on the Euler--Maclaurin
formula.\\

Further, with $G(\tilde{y})$ calculated in the Chebyshev nodes $\tilde{y}_k$ (see \eqref{Tak1}), we compute the integral of
$G(\tilde{y})$ over the interval $[-1,1]$ evaluating the coefficients of expansion of $G(\tilde{y})$ in the polynomial series using the FFT. Then we take their linear combination similar to \eqref{C3a} and finally get the value of $I_{\psi3}$.

To integrate over the subdomain $\Sigma$, we first maps $\Sigma$ on the first quadrant
$\{(\check{y},\check{z}):0\leq\check{y},\check{z}<\infty\}$ using the linear mapping $\eta_2=\check{y}+\eta_{max}$,
$\eta_3=\check{z}+\eta_{max}$ with the unit Jacobian and then proceed as above to write
$$
I_{\Sigma}=\int\limits_{0}^{\infty}\biggl(\int\limits_{0}^{\infty}g(\check{y}+\eta_{max},\check{z}+\eta_{max})d\check{z}\biggr)
d\check{y}=\int\limits_{0}^{\infty}G(\check{y}) d\check{y}.
$$
Further we take $\check{y}_k=\exp\bigl(\dfrac{\pi}{2}\sinh kh\bigr)$, $k=-N_t,...,N_t$ and define $\check{z}(u)$ by the formula $\check{z}(u)=\exp\bigl(\dfrac{\pi}{2}\sinh u\bigr)$ with the Jacobian
$J_T(u)=\dfrac{\pi}{2}\check{z}(u)\cosh(u)$. The integrals $G(\check{y}_{-N_t})$,..., $G(\check{y}_{N_t})$ and finally
$\int\limits_{0}^{\infty}G(\check{y}) d\check{y}$ are calculated by using the relationships presented in  formula \eqref{Tak1}.

\subsection{Feature 3: power law components of the integrand}
\label{feature3}

The integrands of the operators $A(f,\eta)$ and $B(f,\eta)$ could have a singularity near the boundary of
the domain $\Delta_{\eta}$ (see the red lines in Fig.~\ref{fig1}). This singularity is of the form $\eta^{- p}$ and this is due to the formula~(\ref{E10_1}).  It significantly decrease the rate of convergence of the
cubature formulas applied for computing the integrals. Moreover, it does not allow to obtain sufficiently accurate results in the acceptable time.

The following two cases appear:

1. The case of the subdomains
$\Delta_{1,4,5}$,
$\psi_{2,3}$, $\rho_{1,2}$, which are  located along the dashed line in Fig.~\ref{fig2}.  After applying the map $\mathfrak{F}$ to these subdomains we come to the problem of calculating the integrals of the form 
\begin{equation}
\label{sing1}
\lambda(y)=\int_{-1}^1 (z+1+\varepsilon)^q{h}(y,z)dz,~~q \in \mathbb R, \quad  0 \leq \varepsilon\ll 1,
\end{equation}
where ${h}(y,z) \sim 1$ as $z \to -1$ and $q$ is the value which is a priori unknown (see \eqref{EA7}). For the cases when $\varepsilon$ is only 2--3 orders less then unity (this is precisely the case realized in the calculations given in Section~\ref{S5}), the problem of computing the
integral in subdomains located along the dashed line in Fig.~\ref{fig2} can be solved by using the proposed cubature formulas with a moderate number of nodes;

2. The case of the subdomains  $\Delta_{6}$--$\Delta_{9}$, T$_1$--T$_5$, $\psi_1$, $\psi_4$ that have common boundary with the domain $\Delta_{\eta}$ also leads to the problem of integration of the form \eqref{sing1}. However in this case $\varepsilon=0$ and $q=1/2$ that follows from the presence of the square-root singularity in formula \eqref{C1_0} for the kernel $S$ (see \eqref{EA4} for details). In this case the integral can be calculated by using the Gauss--Jacobi quadrature (see \cite{ralston2001first}, $\S$ 4.8.1).
However, this class of quadrature formulas requires to recompute nodes and weights when the number of nodes changes and, as it was shown in
\cite{weideman2007kink} for classical Gauss formula in the case when the integrand has singularities, it is not superior to the Clenshaw--Curtis quadrature which is more simple and
efficient for practical purposes. Therefore, we present the modified Clenshaw--Curtis formula for integrating the functions of the
form $(1\pm z)^qh(y,z)$. 

We denote $I_h(y_k)=\int_{-1}^1(1\pm z)^{q}h(y_k,z)dz$ and use the series $h(y_k,z)\approx\sum\limits_{m=0}^{M}a_{mk}T_m(z)$, $T_m(z)=\cos(m\arccos(z))$. With this, we can write
\begin{equation}\label{C8}
I_h(y_k)\approx I_h^M(y_k)=\sum\limits_{m=0}^{M}a_{mk}\int_{-1}^1(1\pm
z)^{q}T_m(z)dz=\sum\limits_{m=0}^{M}a_{mk} \mathcal{T}_m^q
\end{equation}
and after cumbersome computations we have 
\begin{equation}
\label{SemF}
\begin{aligned}
&~~~~~~\mathcal{T}_m^q=\int_{-1}^1(1\pm z)^{q}T_m(z)dz=\frac{(\pm 1)^m}{2^{1+q}(p^2-m^2)((1+q)^2-m^2)}\times\\
&\times\left\{4^{q+1}q(m^2+q+q^2)+m(1+2q)[-(m+q)G_q^{-m}+(q-m)G_q^m\right\},
\end{aligned}
\end{equation}
where $G_q^m:={}_2F_1(m-q-1,-2q;m-q;-1)$ is the hypergeometric function.
Here ${}_2F_1(a,b;c;z):=\sum\limits_{n=0}^{\infty}\dfrac{(a)_n(b)_nz^n}{(c)_nn!}$,
$(a)_n=(a+1)...(a+n)$ is the Pochhammer symbol.
Notice that for $q = 1/2$ the formula~(\ref{SemF}) reads 
$$
\mathcal{T}_m^{\frac{1}{2}} = 4\sqrt{2}(\pm 1)^m\frac{3-4m^2}{(1-4m^2)(9-4m^2)}.
$$

The following assertion holds
\begin{proposition}
\label{theo2}
If $h(y_k,z)$ in \eqref{C8} is an analytic function with respect to $z$ then the error of approximation of the formula \eqref{C8} decreases exponentially as $M \to \infty$.
\end{proposition}
The proof of the proposition~\ref{theo2} and numerical check of the convergence of \eqref{C8} is given in appendix~3. 

The detailed verification of the cubature formulas for computing the integrals $A(f,\eta)$ and $B(f,\eta)$ is given in appendix~4.

\section{Algorithm for solving the problem \eqref{C1}, \eqref{C2}}

This section is devoted to a numerical algorithm for solving the nonlinear eigenvalue problem \eqref{C1}, \eqref{C2}. To align the form of
equation \eqref{C1}, which includes the first-order differential operator, with the number (two) of boundary conditions in \eqref{C2}, we
differentiate it with respect to  $\eta$:
\begin{equation}
\label{Ap2_1}
\eta\frac{d^2 f}{d\eta^2}+(1+x)\frac{d f}{d\eta}=\frac{1}{b}\biggl(\frac{d A(f,\eta)}{d\eta}+f\frac{d B(f,\eta)}{d\eta}+\frac{d
f}{d\eta}B(f,\eta)\biggr).
\end{equation}
Instead of \eqref{C1}, \eqref{C2}, we study numerically the boundary value problem \eqref{Ap2_1}, \eqref{C2}. Notice that equation \eqref{Ap2_1}
corresponds to equation \eqref{C1} by adding arbitrary constant $C_a$ in \eqref{C1}, and we have to find a solution of \eqref{Ap2_1},
\eqref{C2} which is also solution of the original equation \eqref{C1}, i.e. for which $C_a=0$.
We will take $\eta_{max} \in [10,100]$ in the boundary condition \eqref{C2} and look for $\eta_{min}$ 
from the interval $(0,\eta_{max}/10]$. The bisection method is applied for finding the
value of $\eta_{min}$ such that the following nonlinear equation is satisfied
 \begin{equation}
\label{Ap2_2}
xf(\eta_{min})=\frac{1}{b}(A(f,\eta_{min},x)+f(\eta_{min})B(f,\eta_{min},x)).
\end{equation}
With this and the first boundary condition from \eqref{C2}, we derive that $C_{a}=0$.

To continue, we introduce the operator $\Lambda[f]$ and
rewrite \eqref{Ap2_1} in the form
\begin{equation}
\label{Ap2_3} \Lambda[f] f:= \eta\frac{d^2 f}{d\eta^2}+\biggl(1+x-\frac{B(f,\eta)}{b}\biggr)\frac{d f}{d\eta}-\frac{1}{b}\frac{d
B(f,\eta)}{d\eta}~f=\frac{1}{b}\frac{d A(f,\eta)}{d\eta}.
\end{equation}
We consider the boundary value problem \eqref{Ap2_3}, \eqref{C2} under the constrain \eqref{Ap2_2} which enables us to find $\eta_{min}$.
The two steps for finding the solutions  are suggested. First, the problem will be reduced to a linear
algebra spectral problem. To this end, we linearize the equation to construct an iterative relaxation process and seek an  approximation of the solution on each iteration by interpolation polynomial with Chebyshev nodes. As a result, we define the consecutive approximations of
the exponent $x$ as a solution of the corresponding linear spectral problem. The process is stopped if the
residual of the problem reaches a specified tolerance. The second step consists in applying the relaxation method to
improve the stability of the obtained numerical solution and to reduce the error.

\subsection{Relaxation method }
\label{S4_0}
The linearization of the equation \eqref{Ap2_3} which we apply for the first step, is used in the form suggested in \cite{blokhin2009numerical}. This linearization also   leads to the scheme of the relaxation process of the second step.
We introduce a new variable $\tilde{t}$ which plays the role of time in relaxation method and consider $f=f(\tilde{t},\eta)$. We find a solution of~\eqref{Ap2_3} as the limit of $f(\tilde{t},\eta)$, which satisfies the equation
\begin{equation}
\label{EE4} \mathfrak R_{\tilde{t}} f= \Lambda[f] f-\frac{1}{b}\frac{d A(f,\eta)}{d\eta},
\end{equation}
when $\tilde{t} \rightarrow\infty$. We will consider the following two regularization operators: $\mathfrak
R_{\tilde{t}}=\dfrac{\partial}{\partial \tilde{t}}$ (simple regularization) and
$\mathfrak R_t=(k_1-k_2\Lambda[f])\dfrac{\partial}{\partial \tilde{t}}$
(non-trivial regularization). Here  $k_1 > 0$ and $k_2 > 0$ are  parameters.
The method converges if
$\dfrac{\partial f}{\partial \tilde{t}}\rightarrow0$ as
$\tilde{t}\rightarrow\infty$ and the residual $\varepsilon_S=\|\mathfrak
R_{\tilde{t}} f\|$ for $\tilde{t} \gg 1$ becomes sufficiently small. Here $\|\cdot\|$ denotes the supremum norm. 
With this, we say that
$f(\tilde{t},\eta)$ is an approximate solution of \eqref{Ap2_3} for large $\tilde{t}$.
We implemented the possibility to stop the iterative procedure and to change the regularization of the problem and its parameters during computational process. This enabled us to decrease the residual function (see below).

To organize the iterative process, we introduce the grid over $\tilde{t}$ with the nodes $\tilde{t}_n=\tilde{\tau} n$ and the step $\tilde{\tau}$ and denote
$f^{[n]}=f^{[n]}(\eta)=f(\tilde{t}_n,\eta)$. The time
derivative is approximated by the difference quotient
$\dfrac{\partial f}{\partial \tilde{t}}(\tilde{t}_n,\eta)\approx\dfrac{f^{[n+1]}-f^{[n]}}{\tilde{\tau}}$. For the simple regularization operator, we have
\begin{equation}\label{EE5}
(1-\tilde{\tau}\Lambda[f^{[n]}])f^{[n+1]}=f^{[n]}-\tilde{\tau}\frac{1}{b}\frac{d A(f^{[n]},\eta)}{d\eta}.
\end{equation}
For the non-trivial regularization operator, we get
\begin{equation}
\label{EE6}
\big(k_1-(k_2+ \tilde{\tau})\Lambda[f^{[n]}]\big)f^{[n+1]}=(k_1-k_2\Lambda[f^{[n]}])f^{[n]}- \tilde{\tau}\frac{1}{b}\frac{d A(f^{[n]},\eta)}{d\eta}.
\end{equation}
{\em A priori} analysis of the convergence of the relaxation method  described for the problem under consideration is unavailable.
However, experimentally we find the parameters $\tilde{\tau}$, $k_1$, $k_2$ that provide  convergence of the method.
Thus, it remains to do the following things: approximation of $f(\eta)$ and its derivatives; the statement of the spectral problem; realization of the algorithm for refinement of the  solutions obtained by the spectral problem. 

\subsection{Approximation of $f(\eta)$ and its derivatives}
\label{S4_1}
The description of the approach contains rather cumbersome derivations. We refer the reader to appendix~5 for the detail. The outcome of these derivations are the formulas of matrix approximation of the differentiation operation:
\begin{equation}
\label{AprMatr} \mathbf{F}_{\eta}\approx\mathcal A\mathbf{F},\ \ \ \mathbf{F}_{\eta\eta}\approx\mathcal B\mathbf{F}.
\end{equation}
Here $\mathbf{F}$, $\mathbf{F}_{\eta}$, $\mathbf{F}_{\eta\eta}$ are the vectors of the values of the solution $f(\eta)$ and its derivatives written in $N+2$ nodes $e_0,...,e_{N+1}$ that are $N$ nodes $e_1,...,e_{N}$ of Chebyshev mesh specified on the interval $[\eta_{min},\eta_{max}]$ supplemented with its end points $e_0=\eta_{min}$, $e_{N+1}=\eta_{max}$, and $\mathcal A$ and $\mathcal B$ are the $(N+2)\times (N+2)$ matrices containing the coefficients of the derivatives of interpolation polynomials between these nodes (see appendix~5).
\begin{remark}
Notice that here we use interpolation polynomial for approximating the solution $f(\eta)$ and the only information that we need to keep in memory is the value of $f$ in the nodes of interpolation.
To calculate $f(\eta)$ at arbitrary point $\eta\in[\eta_{min},\eta_{max}]$ by $f(e_j)$, $j=1,...,N$, we use the algorithm which is based on the barycentric representation of the interpolation polynomial~\cite{salzer1972lagrangian}.
It requires $O(N)$ operations and can be vectorized. We also explore this algorithm for computing the integrands of the operators $A(f,\eta)$, $B(f,\eta)$ at the nodes of integration located in subdomains shown in Fig.~\ref{fig2}. The barycentric representation can be generalized for the case of rational interpolations, see \cite{tee2006rational}. We also tried this kind of approximation in our problem, but it didn't give essential advantages.
\end{remark}

\subsection{Spectral problem}
\label{S4_2}

We use  the collocation method with the nodes $e_0,...,e_{N+1}$ to approximate the regularization operators applied for the equation \eqref{Ap2_3}. Let $\mathbf{F}^{[n+1]}$ be the vector containing the values of $f$ at the nodes $e_j$ obtained on the $(n+1)$st iteration.

For the simple regularization, taking into account \eqref{AprMatr}, we get the following equation for $\mathbf{F}^{[n+1]}$
\begin{equation}
\label{apEq1}
(E-\tilde{\tau}\widetilde{\Lambda}_n)\mathbf{F}^{[n+1]}=\mathbf{F}^{[n]}-\frac{\tilde{\tau}}{b^{[n]}}
\text{diag}(A'_{\eta}(\mathbf{F}^{[n]},\mathbf{e})),
\end{equation}
where $\widetilde{\Lambda}_n$ is the matrix approximation of the operator $\Lambda[f^{[n]}]$:
$$
\widetilde{\Lambda}_n=\text{diag}(\mathbf{e})\mathcal{B}+\text{diag}\biggl(1+x^{[n+1]}-\frac{1}{b^{[n]}}B(\mathbf{F}^{[n]},\mathbf{e})\biggr)\mathcal{A}-\frac{1}{b^{[n]}}\text{diag}(B'_{\eta}(\mathbf{F}^{[n]},\mathbf{e})).
$$
Here $A(\mathbf{F}^{[n]},\mathbf{e})$ and
$B(\mathbf{F}^{[n]},\mathbf{e})$ are the vectors of values of
$A(f^{[n]}, \eta)$ and $B(f^{[n]},\eta)$ at the points $\mathbf{e}$,
$\text{diag}(\mathbf{e})$ means a diagonal matrix with the
components of the vector $\mathbf{e}=(e_0,...,e_{N+1})$ located on the diagonal, $E$ is the unit
matrix. The vectors $A'_{\eta}(\mathbf{F}^{[n]},\mathbf{e})$ and $B'_{\eta}(\mathbf{F}^{[n]},\mathbf{e})$ contain the derivatives of $A(f,\eta)$ and $B(f,\eta)$ at the nodes $e_j$, $x^{[n]}$ is the $n$th approximation of $x^*$, $b^{[n]}=\dfrac{1}{2(x^{[n]}-1)}$.

For the non-trivial regularization \eqref{EE6} of equation \eqref{Ap2_3}, we have
 \begin{equation}
\label{apEq2}
\big[k_1E-(k_2+\tilde{\tau})\widetilde{\Lambda}_n\big]\mathbf{F}^{[n+1]}=k_1\mathbf{F}^{[n]}-k_2\widetilde{\Lambda}_n\mathbf{F}^{[n]}-
\frac{\tilde{\tau}}{b^{[n]}}\text{diag}(A'_{\eta}(\mathbf{F}^{[n]},\mathbf{e})).
\end{equation}

Relations \eqref{apEq1}, \eqref{apEq2} enable to pose the following spectral problem.

To find $\mathbf{F}^{[n+1]}$ and the spectral parameter $\lambda^{[n+1]}$ which satisfy 
\begin{equation}\label{spec}
(\mathfrak{M}^{[n]}-\lambda^{[n+1]}E)\mathbf{F}^{[n+1]}=0. 
\end{equation}
The approximate value of $x^*$ is determined by 
\begin{equation}\label{speca}
x^{[n+1]}=-\lambda^{[n+1]} - 1.
\end{equation}

For the simple regularization by virtue of \eqref{apEq1}, we have
\begin{equation}\label{simpR}
\begin{aligned}
&~~~~~~~~~~~~~~~~~~\mathfrak{M}^{[n]}=\biggl(\text{diag}\biggl(1-\frac{B(\mathbf{F}^{[n]},\mathbf{e})}
{b^{[n]}(1+x^{[n]})}\biggr)\mathcal A - \tilde{\tau} E\biggr)^{-1}\\
   &\left\{\text{diag}(\mathbf{e})\mathcal A-\text{diag}\biggr(- \tilde{\tau}(1+x^{[n]})+\frac{1}{b^{[n]}}\biggl[B'_{\eta}(\mathbf{F}^{[n]},\mathbf{e})+
   \frac{A'_{\eta}(\mathbf{F}^{[n]},\mathbf{e})}{\mathbf{F}^{[n]}}\biggr]\biggr)\right\}.
    \end{aligned}
\end{equation}
Here $\dfrac{A'_{\eta}(\mathbf{F}^{[n]},\mathbf{e})}{\mathbf{F}^{[n]}}$ denotes the component-wise division of the
vector $A'_{\eta}(\mathbf{F}^{[n]},\mathbf{e})$ by the vector
$\mathbf{F}^{[n]}$.

For the non-trivial regularization in accordance with \eqref{apEq2} the operator reads
\begin{equation}\label{nontR}
\mathfrak{M}^{[n]}=
\biggl(\text{diag}\biggl(1-\frac{B(\mathbf{F}^{[n]},\mathbf{e})}{b^{[n]}(1+x^{[n]})}\biggr)\mathcal A+\frac{k_1}{\tilde{\tau}}E-\frac{k_2}{\tilde{\tau}}\mathcal B\biggr)^{-1}\mathfrak{B}^{[n]},
\end{equation}
where
$$
\mathfrak{B}^{[n]}=\text{diag}\biggl(\mathbf{e}+(1+x^{[n]})\frac{k_2}{\tilde{\tau}}\biggr)\tilde{\mathcal
B}-\text{diag}\biggr(\frac{(1+x^{[n]})k_1}{\tilde{\tau}}+
\frac{1}{b^{[n]}}\biggl[B'_{\eta}(\mathbf{F}^{[n]},\mathbf{e})+\frac{A'_{\eta}(\mathbf{F}^{[n]},
\mathbf{e})}{\mathbf{F}^{[n]}}\biggr]\biggr).
$$

To proceed, we use the eigenvalue $x^{[n]}$ and the eigenvector $\mathbf{F}^{[n]}$ to calculate the elements of
matrix $\mathfrak{M}^{[n]}$. Then, we define $x^{[n + 1]}$ and $\mathbf{F}^{[n + 1]}$ from \eqref{spec} such that
the following conditions are satisfied:
\begin{itemize}
\item[1.] The eigenvalue $x^{[n+1]}$ obtained by the spectral parameter $\lambda^{[n+1]}$ is real number and $x^{[n+1]}\in(1,1.5)$.
\item[2.] $\mathbf{F}^{[n+1]}$ has only positive bounded components.
\item[3.] The relative value $\dfrac{\max|\mathbf{F}^{[n+1]}-\mathbf{F}^{[n]}|}{\tilde{\tau} \max|\mathbf{F}^{[n]}|}$ is small enough and the
    difference $|x^{[n+1]}-x^{[n]}|$ is small enough too.
\end{itemize}
Then  $\mathbf{F}^{[n+1]}$ and the corresponding eigenvalue $x^{[n+1]}$ are determined by the criterion of minimum of the relative
residual function, see \eqref{Rr}. Using $\mathbf{F}^{[n+1]}$ and $x^{[n+1]}$, we calculate the matrix $\mathfrak{M}^{[n+1]}$ and continue the iterations. The iteration
process is stopped when the maximum value of the relative residual $\|\mathbf{R}_r(\eta)\|$ becomes small enough. Notice that on each
iteration we specify $\eta_{min}$ by the formula \eqref{Ap2_2}. In Table~\ref{tab3} the range of values $\eta_{max}$ and the ranges of
initial values $\eta_{min}$ and $x$ are presented.

This relative residual function and its discrete counterpart are as follows:
\begin{equation}
\label{Rr}
R_r(\eta)=\left\{
\begin{aligned}
&R_r^0,~0\leq\eta<\eta_{min},\\
&R_r^c,~\eta_{min}\leq\eta\leq\eta_{max},\\
&R_r^{\infty},~\eta>\eta_{max}.
\end{aligned}
\right.~~~\mathbf{R}_r=\left\{
\begin{aligned}
&\mathbf{R}_r^0,~0\leq\eta<\eta_{min},\\
&\mathbf{R}_r^c,~\eta_{min}\leq\eta\leq\eta_{max},\\
&\mathbf{R}_r^{\infty},~\eta>\eta_{max},
\end{aligned}
\right.
\end{equation}
where
\begin{equation}\label{relRes}
\begin{split}
& R_r^0(\eta)=R_r^0(x, C_0, \eta)=\dfrac{\big|xbC_0-\bigl(A(C_0,\eta)+C_0B(C_0,\eta)\bigr)\big|}{\max\left\{xbC_0,~\big|A(C_0,\eta)\big|,~\big|C_0B(C_0,\eta)\big|\right\}},~C_0=f(\eta_{min}),\\
&R_r^c(\eta)=R_r^c(x, f, \eta)=\dfrac{\bigg|xf+\eta \dfrac{d
f}{d\eta}-\dfrac{1}{b}\biggl(A(f,\eta)+fB(f,\eta)\biggr)\bigg|}{\max\left\{\big|xf(\eta)\big|,~\bigg|\eta \dfrac{d
f}{d\eta}\bigg|,~\bigg|\dfrac{A(f,\eta)}{b}\bigg|,~\bigg|f\dfrac{B(f,\eta)}{b}\bigg|\right\}},\\
&R_r^{\infty}(\eta)=R_r^{\infty}(x, C_{\infty}, \eta)=\dfrac{\big|A(C_{\infty}\eta^{-x},\eta)+C_{\infty}\eta^{-x}B(C_{\infty}\eta^{-x},\eta)\big|}{\max\left\{\dfrac{xbC_{\infty}}{\eta^x},~\bigg|A\bigg(\dfrac{C_{\infty}}{\eta^x},\eta\bigg)\bigg|,~\bigg|\dfrac{C_{\infty}}{\eta^x}
B\bigg(\dfrac{C_{\infty}}{\eta^x},\eta\bigg)\bigg|\right\}}.
\end{split}
\end{equation}
We denoted the values of $R_r(\eta)$ obtained on the $n$th iteration at the nodes $\mathbf{e}$ by the vector $\mathbf{R}_r^c=R_r^c(x^{[n]}, \mathbf{F}^{[n]},
\mathbf{e})$:
\begin{equation}\label{relRes2}
\mathbf{R}_r^c=\dfrac{\bigg|x^{[n]}\mathbf{F}^{[n]}+\text{diag}(\mathbf{e})\mathcal{A}\mathbf{F}^{[n]}-\dfrac{1}{b}\biggl(A(\mathbf{F}^{[n]},\mathbf{e})+\mathbf{F}^{[n]}\cdot
B(\mathbf{F}^{[n]},\mathbf{e})\biggr)\bigg|}{\max_{c}\left\{\big|x^{[n]}\mathbf{F}^{[n]}\big|,~\bigg|\text{diag}(\mathbf{e})\mathcal{A}\mathbf{F}^{[n]}\bigg|,~\bigg|\dfrac{A(\mathbf{F}^{[n]},\mathbf{e})}{b}\bigg|,~\bigg|\mathbf{F}^{[n]}\cdot
\dfrac{B(\mathbf{F}^{[n]},\mathbf{e})}{b}\bigg|\right\}}.
\end{equation}
Here the symbol $\cdot$ denotes the component-wise product, the division is also the component-wise operation, $\max_{c}$ denotes the component-wise maximum,
$|\mathbf{v}|$ denotes the vector which is composed of absolute values of the components of $\mathbf{v}$. 
We also computed the values of residual function outside the segment $[\eta_{min},\eta_{max}]$
\begin{equation}\label{relRes0}
\mathbf{R}_r^0=\mathbf{R}_r^c(x^{[n]}, C_0^{[n]},
\boldsymbol{\zeta})=\dfrac{\big|x^{[n]}bC_0^{[n]}-\bigl(A(C_0^{[n]},\boldsymbol{\zeta})+C_0^{[n]}
B(C_0^{[n]},\boldsymbol{\zeta})\bigr)\big|}{\max_{c}\left\{x^{[n]}bC_0^{[n]},~\big|A(C_0^{[n]},\boldsymbol{\zeta})\big|,~\big|C_0^{[n]}B(C_0^{[n]},\boldsymbol{\zeta})\big|\right\}},
\end{equation}
\begin{equation}\label{relResInf}
\mathbf{R}_r^{\infty}=\mathbf{R}_r^{\infty}(x^{[n]}, C_{\infty}^{[n]},
\boldsymbol{\xi})=\dfrac{\big|\bigl(A(\boldsymbol{\vartheta}^{[n]},\boldsymbol{\xi})+\boldsymbol{\vartheta}^{[n]}\cdot 
B(\boldsymbol{\vartheta}^{[n]},\boldsymbol{\xi})\bigr)\big|}{\max_{c}\left\{b\boldsymbol{\vartheta}^{[n]},~\big|A(\boldsymbol{\vartheta}^{[n]},\boldsymbol{\xi})\big|,~\big|\boldsymbol{\vartheta}^{[n]}\cdot B(\boldsymbol{\vartheta}^{[n]},\boldsymbol{\xi})\big|\right\}}.
\end{equation}
Here $C_0^{[n]}$ is the value of $f^{[n]}(\eta_{min})$ computed on the $n$th iteration, $C_{\infty}^{[n]}$ can be computed by \eqref{Cinf}, $\boldsymbol{\zeta}=(\zeta_1,...,\zeta_K)$, $\boldsymbol{\xi}=(\xi_1,...,\xi_M)$, $\boldsymbol{\vartheta}^{[n]}=(\vartheta_1^{[n]},...,\vartheta_M^{[n]})$ are the vectors with components laying outside the segment $[\eta_{min},\eta_{max}]$:
$$
\zeta_k=\frac{k-1}{K}\eta_{min},~~~\xi_m=\eta_{max}+\frac{4m}{M}\eta_{max},~~~\vartheta_m^{[n]}=C_{\infty}^{[n]}\xi_m^{-x^{[n]}},~~~k=1,...,K,~m=1,...,M.
$$
In the order to compute the components of vectors $A(C_0^{[n]},\boldsymbol{\zeta})$, $B(C_0^{[n]},\boldsymbol{\zeta})$, $A(\boldsymbol{\vartheta}^{[n]},\boldsymbol{\xi})$, $B(\boldsymbol{\vartheta}^{[n]},\boldsymbol{\xi})$ we considered these integrals as the difference of two integrals over first quadrant and over triangle located in the lower left corner of this quadrant (see the white triangle in Fig.~\ref{fig1}). For computing the integral over first quadrant, the double exponential formulas were applied similarly as in the case of integration over the domain $\Sigma$, see section~\ref{F2}. For computing the integral over triangle the transformations shown in Fig.~\ref{fig2_2} were done, and the values of $f_c(\eta)=f(\eta_2+\eta_3-\eta)$ were set equal to zero in this triangle. This method of computing the integrals does not account for the smoothness of the solution and therefore it is not so fast and accurate as the method described in section~\ref{intComp}. However, it was used only for computing residual function in few points (usually we set $K=M=5$), and therefore it doesn't reduce the high speed of computations.

The initial function for the iterative process is taken of the form
\begin{equation}
\label{initF}
f_0(\eta)=\left\{
\begin{aligned}
&C_0,~0\leq\eta<\eta_{min},\\
&-p_I(\eta),~\eta_{min}\leq\eta\leq\eta_{max},\\
&\eta^{-x_I},~\eta>\eta_{max}.
\end{aligned}
\right.
\end{equation}
Here $p_I(\eta)$ is a polynomial satisfying the conditions: (a) $p_I(\eta_{min})=C_0$ with some positive constant $C_0$, (b) the
derivatives $p_I^{(k)}$ at the point $\eta_{min}$ equal zero for $k=1,...,N_{min}$, (c) $p_I(\eta_{max})=\eta_{max}^{-x_I}$ and (d) the
derivatives $p_I^{(k)}$ at the point $\eta_{max}$ equal $-x_I(-x_I-1)...(-x_I-k+1)\eta_{max}^{-x_I-k}$, $k=1,...,N_{max}$. We can easily
determine the unique polynomial $p_I(\eta)$ of the power $N_{min}+N_{max}+1$ which satisfy the conditions (a)--(d). Substituting $p_I$ into
the formula \eqref{initF}, we conclude that $f_0(\eta)$ is the $N_{min}$ times continuously differentiable function at the point $\eta_{min}$ and
the $N_{max}$ times continuously differentiable function at the point $\eta_{max}$. The parameters $C_0$, $x_I$, $\eta_{min}$, $\eta_{max}$,
$N_{min}$, $N_{max}$ used for the computations are specified by the Table~\ref{tab3}.

\subsection{Refinement of the spectrum }
\label{S4_3}

Solving the spectral problems on each step of the procedure described above with the different parameters $C_0$, $x_I$, $\eta_{min}$, $\eta_{max}$, $N_{min}$, $N_{max}$, we found a set of functions that approximate steady states of
equation~(\ref{EE4}). Further we apply the relaxation method  to refine the obtained steady states.
We use again \eqref{apEq1}, \eqref{apEq2} to organize the iteration process. We take the initial data $\mathbf{F}^{[0]}$ and $\lambda^{[0]}$ equal to the spectrum and the spectral parameter determined on the previous step.
Then $\mathbf{F}^{[n + 1]}$, $n=1,2,...$  is calculated by the inversion of the corresponding matrices in the LHS of \eqref{apEq1}, \eqref{apEq2} without solving the spectral problem. 
Due to the special structure of the matrices $\mathcal{A}$ and $\mathcal{B}$, the functions obtained on each iteration automatically satisfy the boundary conditions \eqref{C2}, see appendix~5. Notice that to vanish the constant $C_a$ on each iteration, we recalculate the exponent $x$ (derived in the previous section) by solving equation \eqref{Ap2_2} with respect to $x$ exploiting the bisection method. As we will see in the next section, this procedure leads to improvement of the spectra obtained.

\section{Results }
\label{S5}

To solve the boundary value problem \eqref{Ap2_3}, \eqref{C2}, we suggested an algorithm  which was realized in two steps. The first step consists in solving the spectral problem \eqref{spec} by an iterative process with the initial data \eqref{initF}. The parameters of the spectral problem and the ranges of their values are specified in Table~\ref{tab3}.
We select 7 suitable numerical solutions of the problem under consideration satisfying the conditions 1--3 formulated in
section~\ref{S4_2} which provide the relative residual \eqref{Rr} less than 12\%.
We group them into three sets. The first set contains the solutions with $x<1.2$, the second set is spectra with $1.2\leq x\leq 1.23$ and the third one contains spectra with $x>1.23$. Table~\ref{tab4} presents the parameters of all the spectra. We plot the spectra from the first set in Fig.~\ref{fig7}, panel (a) with the respective relative residual functions $R_r(\eta)$  presented on the Panel (b). The spectra and the residuals for the sets 2 and 3 look qualitatively similar, so we have chosen not to present the respective plots.
\begin{table}[!hbt]
\caption{Values of the parameters used in computations}
\label{tab3}
\begin{center}
   \begin{tabular}{c|l|c}  \hline
Parameter   & Description & Range of values  \\[1ex]
    \hline
$\eta_{min}$ & The left boundary of the interval of problem & 1 --
10 \\
$\eta_{max}$ & The right boundary of the interval of problem & 10 --
100 \\
$N$ & The number of interpolation nodes & 41 -- 81 \\
$\tilde\tau$ & The step of time grid & 0.005 -- 1 \\
$x_I$ & The initial value of the parameter $x$ & 1.1 -- 1.5 \\
$k_1$ $k_2$ & The parameters of non-trivial regularization & 1 -- 100\\
$C_0$ & The value of the initial function \eqref{initF}  & 0.001 -- 10 \\
 & at the point $\eta=\eta_{min}$ & \\
$N_{min}$, $N_{max}$ & The number of continuous derivatives of & 1--5\\
 & the initial function at points $\eta_{min}$, $\eta_{max}$\\[1ex]
\hline
\end{tabular}
\end{center}
\end{table}
\begin{figure}[ht]
\centering
\includegraphics[scale=0.37]{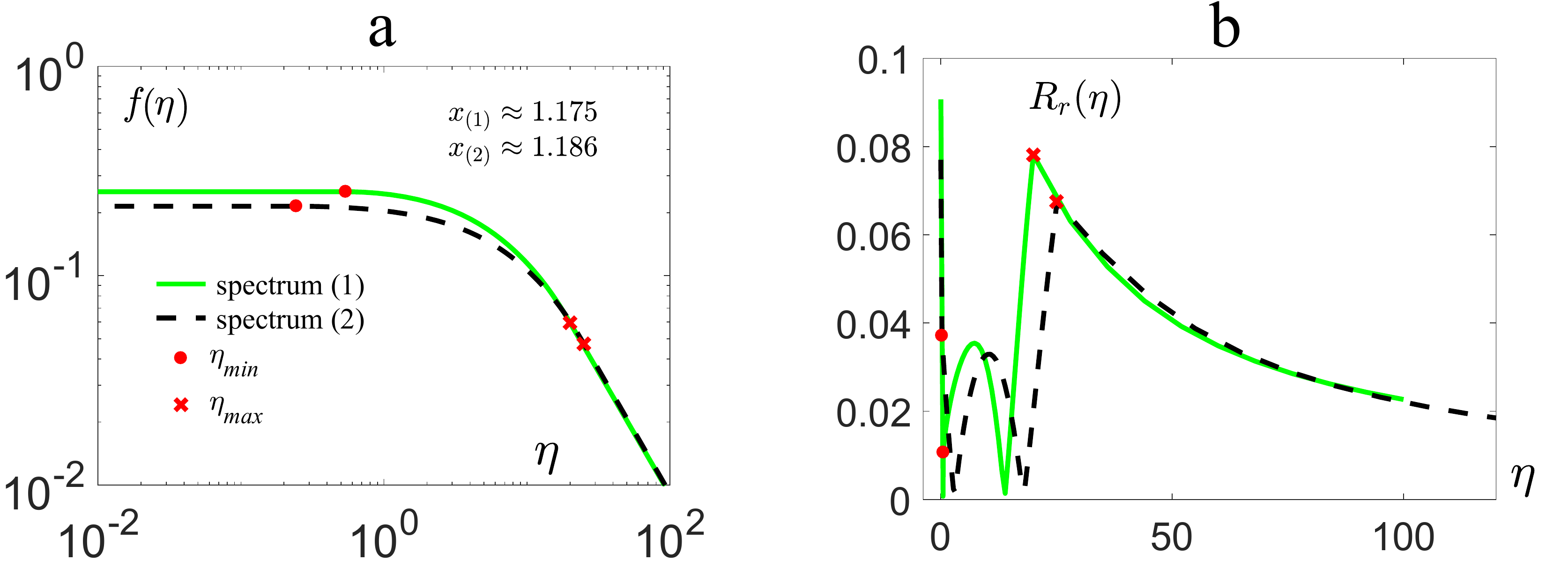}
\caption{(a) Two spectra with the values of $x <1.2$ in the log-log scale. (b) The relative residual function for
$x_{(1)}$, $x_{(2)}$}\label{fig7}
\end{figure}

\begin{table}[!hbt]
\caption{Parameters of the calculated spectra }
\label{tab4}
\begin{center}
  \begin{tabular}{c|cc|ccc|cc}  \hline
  & \multicolumn{7}{|c}{Value of the parameter for the spectrum}\\[1ex]
  Parameter & \multicolumn{2}{c|}{set 1} & \multicolumn{3}{c}{set 2} & \multicolumn{2}{|c}{set 3} \\
  & (1)&(2)&(3)&(4)&(5)&(6)&(7)\\[1ex]
    \hline
$\eta_{min}$ & 0.5404 & 0.2438 & 0.225 & 0.745 & 0.2211 & 0.8277 & 0.524 \\
$\eta_{max}$ & 20 & 25 & 23 & 20& 20 & 25 & 22 \\
$x$ & 1.175 & 1.186 & 1.213 & 1.227 & 1.215 & 1.252 & 1.295 \\
$\|R_r\|$ & 9.07\% & 7.7\% & 8.21\% & 8.97\% & 11.58\% & 11.45\% & 9.65\% \\[1ex]
\hline
\end{tabular}
\end{center}
\end{table}

The residual function usually (but not always) attains maximum value in the vicinity of the point $\eta=0$. It is in this vicinity that  steep gradients and numerous oscillations of the residual function can be observed. 

By applying the second relaxation step described in section \ref{S4_3}  to all of the selected spectra, we have got more accurate results. The least errors were achieved by using spectra 1 and 2 (set 1) as the initial data for this relaxation method.
The results of calculations are shown in Fig.~\ref{fig11} together with the relative residual functions. 
The best result is demonstrated by the refined spectrum 2 with $\|R_r(\eta)\|\approx 4.69\%$. The refined spectrum 1 leads to $\|R_r(\eta)\|\approx 6.11\%$.
\begin{figure}[ht]
\centering
\includegraphics[scale=0.37]{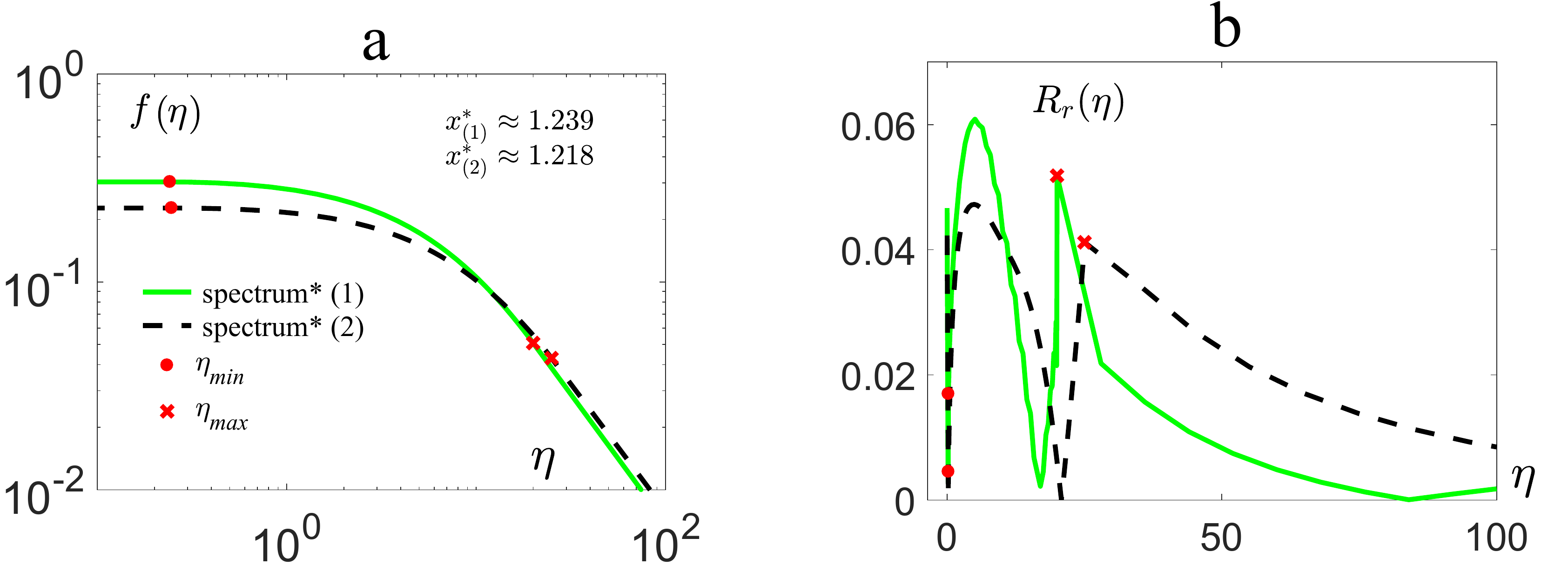}
\caption{(a) The refined spectra 1 and 2 (set 1). (b) The relative residual function for the refined spectra}\label{fig11}
\end{figure}

\begin{color}{black}
We emphasize that the relative residual function measures the accuracy with which the equation for the self-similar function $f(\eta)$ is satisfied
rather than the deviation of $f(\eta)$ from the true solution. We have no means to find the latter in a reliable way and, in particular, we have no rigorous procedure to find the error  the obtained anomalous exponents $x^*$. However, given the fact that our two best solutions satisfy the equation with similar accuracy, we believe that the difference of values of $x^*$ for these solutions gives an indication about the accuracy in finding $x^*$. This leads to an estimate that the found values of $x^*$ are accurate within about 2\%.

\end{color}

\section{Conclusions and discussion }

In this paper, we have studied the self-similar solution to the kinetic equation~(\ref{E1}) which arises in modelling of weak wave turbulence in Bose--Einstein condensates. The self-similarity that arises is of the second kind: it describes a blow-up of the spectrum at the zero frequency in a finite time.

\begin{color}{black}
Let us notice that equation~(\ref{E1}) has been derived under the assumption that the periods of waves $T=2\pi/\omega$ are all much smaller than nonlinear time scale $\tau_{nl}$, which can be estimated as $\tau_{nl} \sim n_{\omega}(t)/\dot{n}_{\omega}(t)$, where $\dot{n}_{\omega}(t)$ is the time derivative of the spectrum.
Decrease of the nonlinear time scale, which inevitably happens in the vicinity of blow-up time $t^*$, leads to the failure of the condition of applicability of the kinetic equation, $\tau_{nl} \gg T=2\pi/\omega$, first at low values of $\omega$, that corresponds to low values  of $\eta$. More precisely, using the definition of self-similar variable $\eta=\omega\tau^{-b}$ and of the solution $f(\eta)=\tau^an_{\omega}(t)$, $\tau=t^*-t$, taking into account that $f_{\eta}(\eta)\to 0$ as $\eta\to 0$ (see section~\ref{NI}), for small $\eta$ we arrive at
\begin{equation}
\label{dop9}
\tau_{nl} \sim \frac{n_{\omega}}{\dot{n}_{\omega}}\bigg|_{\omega\to 0}=\frac{\tau^{-a}f(\eta)}{a\tau^{-a-1}f(\eta)+f_{\eta}(\eta)\dot{\eta}}\bigg|_{\eta\to 0}=\frac{\tau}{a}.
\end{equation}

Hence, $\tau_{nl} \gg 2\pi/\omega$ gives the condition for the scale $\omega\gg 2\pi a/(t^*-t)$ at which our considerations remain valid. Here $a$ is related with anomalous exponent by the expression $a=x^*/[2(x^*-1)]$.

Notice, that the self-similar solution of~(\ref{E1}) admits the group of scaling transformations~\eqref{C1_3}. However, if in the last derivations we make the change $f=\frac{1}{C}\hat{f}(C\eta)$ with arbitrary positive constant $C$, it will lead to exactly the same condition for $\omega$.
\end{color}

Let us now discuss the features of our work related to development of the numerical algorithms for finding the collision integral in equation~(\ref{E1}) and the self-similar solutions on the one hand, and the mathematical and physical meanings of the obtained results on the other.

First, we have developed and tested a new solver for finding the collision integral of equation~(\ref{E1}) which takes into account the singularities of the integrand function and its derivatives. It consists of splitting the integration area into bounded and unbounded subdomains of various shapes, mapping these subdomains  onto a square or a half-stripe, and applying Chebyshev approximations whose coefficients are found via a spectral method. This method has been benchmarked and it has shown superior results in terms of an excellent accuracy and convergence with respect to the mesh refinements.  The singularities arise e.g. due to the minimum function and the square-root components in the integrand's kernel. The other type of quasi-singularities are due to a peaked or/and  oscillatory/sporadic behavior of the intermediate iterations of the spectral shapes. On a qualitative level, they can be attributed to singularities of solutions at low $\eta$'s when $C\ne0$ in equation~(\ref{E10_1}) due to the fact that $x$ (or/and the other parameters) has not converged to the ``right" value yet. Excellent ability of the newly developed solver makes it a very promising tool for solving the time-evolving solutions beyond the self-similar ansatz. Work is underway on testing this solver in such a capacity.

The second part of our work on finding the self-similar solution is the development of the two-step iterative relaxation process. Despite of our great effort and a sophisticated algorithms that we have developed,    making the relaxation process converge turned out to be a notoriously difficult task. Systematic theoretical study of convergence is practically impossible due to the non-linearity of the problem. The stiffness or/and instability of the problem is probably  worsened by the tendency of the intermediate approximations to form singularities, as we mentioned above. Our state of the art result is a solution with a rather modest accuracy $\sim 4.7\%$ giving the asymptotic power law with exponent $x^*\approx 1.22$, \begin{color}{black} and as it is shown in the section~\ref{S5}, the absolute error of this value is about $0.008$. The value of the exponent of our solution\end{color} is quite close to the values obtained previously by the numerical simulations of spectra evolving out of initial data: 
$x^* \approx 1.24$  in \cite{semikoz1995kinetics,semikoz1997condensation} and $x^* \approx 1.2345$
in  \cite{lacaze2001dynamical,connaughton2004kinetic}. \begin{color}{black}We cannot prove rigorously that  one\end{color} of these values is closer to the true solution, even though it is quite obvious that one should not trust more than two digits after the dot. The previous works have not reported on the error in finding $x^*$, but the compensated plots of the spectra and their derivatives in $\log$--$\log$ scale presented in \cite{semikoz1995kinetics,semikoz1997condensation} point at $10-15\%$ uncertainty in determination of $x^*$. \begin{color}{black}As it was mentioned above in our computations this uncertainly is about  $2\%$.\end{color}

Further, no attempt has been previously made to collapse the spectra at different
times onto a single shape curve by a self-similar transformation, but it is quite clear that the error of such a fit would not be possible to reduce to any reasonable values (perhaps not even less than 100\%) over a wide range of scales. This is because the times at which the spectrum is expected to be self-similar is limited to the moments close to $t^*$, and the frequencies -- far enough from the typical frequencies in the initial data. \begin{color}{black}Thus, the authors of \cite{lacaze2001dynamical,connaughton2004kinetic,semikoz1995kinetics,semikoz1997condensation} could not draw any reliable conclusions with respect to the existence of self-similar solutions and their properties based on only the solutions of the evolution equations. Moreover, in these papers no attempts to estimate the error of the obtained numerical solutions were done.  Careful analysis  of this issue needs a good measure for the error estimation; the latter was provided in our study as the relative residual function \eqref{Rr}. It penalizes the maximum of the numerical error over the whole positive values of the self-similar variable. Applying the designed methods and the error estimate, we can state with the high degree of confidence that the self-similar solution exists, and moreover, we can predict its behaviour with the known accuracy.\end{color}

Taking these facts into account, we conclude that the $4.7\%$ accuracy found for the self-similar shape is a rather good result. It puts the (previously hypothetical) self-similar ansatz of the equation~(\ref{E1}) onto a firmer mathematical ground by finding an approximate self-similar solution with a controlled margin of accuracy. This, of course, still leaves unproven the existence of the exact self-similar solution of the second kind. Considering that this solution is of a blow-up type, it allows to make a connection to the mathematical results of \cite{escobedo2014blow} where a quantum version of 
the equation~(\ref{E1}) was proven to have blow-up solutions if the initial data is concentrated at sufficiently low frequencies. Our solution may shed light on what such blowing-up solutions may typically look like, and this is an interesting question for future study.

Physically, the self-similar solutions of the second kind of the equation~(\ref{E1}) describe a non-equilibrium Bose--Einstein condensation, and the ``nonlinear eigenvalue" $x^*$ determines the rate at which the condensation occurs.
Indeed, at $\omega=0$ we can state that $n_0(t) = \dfrac{f(0)}{(t^*-t)^{a}}$ with $a=\dfrac{x^*}{2(x^*-1)} \approx 2.77$. Question arises: what behavior do we expect for $t>t^*$?  The authors of \cite{semikoz1995kinetics,semikoz1997condensation} argued that there will be transition to a thermodynamic energy equipartition rather that the KZ spectrum.
 On the other hand, the unbounded concentration of the spectrum at the low frequencies will inevitably invalidate the applicability of the kinetic equation~(\ref{E1}), and strong turbulence with vortices of hydrodynamic type will appear \cite{nazarenko2011wave}. Further, if the mode $\omega=0$ is damped then, possibly, the KZ scenario becomes relevant for $t>t^*$. Such post blow-up dynamics and its dependence on the boundary condition at zero frequency deserves further investigation.   

\begin{color}{black}
We would like to finish the discussion by the remark that the fast and accurate method for evaluation of the collision integral that we have developed in the present paper has important applications beyond the problem of finding the self-similar spectra for the problem of Bose--Einstein condensation.
Firstly, this method turns out to be the crucial  for a careful comparative analysis of the original Gross--Pitaevskii model and weak turbulence theory based on the kinetic equation. In our ongoing studies we are solving the non-stationary  kinetic equation and use the designed methods to compute the time-dependent spectra. We further use the obtained spectra for finding  the probability density functions and their cumulants, and we compare the results with the results of direct numerical simulation of the Gross--Pitaevskii equation.

Second, our approach and methods can be further  continued and extended to  self-similar setups for kinetic equations in many other applications. For example, doing minor changes in the described method we can consider the case of 2D condensation, the gravitational waves in early universe (see \cite{galtier2017turbulence}), and turbulence in self-gravitating Bose gas (see \cite{skipp2020wave}). These applications are interesting projects for future research.
\end{color}

\section*{Appendix 1. Analysis of convergence of the collision integral}\label{A1}

Let us study convergence of the collision integral in the RHS of~(\ref{E1}) on a   power-law spectrum
 $n_{\omega}=\omega^{-x}$,~$x>0$.
With this, the collision integral  reads
\begin{eqnarray}
I =\omega^{-1/2} \int S\cdot
(\omega\omega_{1}\omega_{2}\omega_{3})^{-x}
\left({\omega}^{x} + \omega_{1}^{x} - \omega_{2}^{x} - \omega_{3}^{x}\right)\delta(\omega + \omega_1 - \omega_2
- \omega_3)d\omega_1d\omega_2d\omega_3.
\end{eqnarray}

First we consider  region (a): $\omega_1 \ll \omega$
and $\omega_2,~\omega_3 \sim \omega $.
Then $S = \omega_1^{1/2}$ and the respective contribution to the integral is
$I_a \sim \int\omega_{1}^{1/2 - x}d\omega_1.$
This integral converges iff $x < 3/2$.

Now let us consider  case (b): $\omega_1 = \alpha\omega_2 \ll \omega \approx \omega_3$, $\alpha \sim 1, \alpha <1$.
 Then $S = \omega_1^{1/2}$ and
 we can write
\begin{equation}\label{E5}
I_b \sim \int\omega_{1}^{1/2 - 2x}\left(\omega^x + \omega_1^x - \omega_2^x - (\omega + \omega_1 - \omega_2)^x\right)d\omega_1d\omega_2.
\end{equation}    
Taylor expanding $(\omega + \omega_1 - \omega_2)^x$ and assuming that $x \geq 1$ we get after algebraic manipulation that 
$I_b \sim \int\omega_{1}^{5/2 - 2x}d\omega_1d\phi.$
Here, we have rewritten the area element  using the polar coordinates, $d\omega_1d\omega_2 = \rho d\phi d\rho$ where $\rho = \omega_1\sqrt{1 + \alpha^{-2}}$ with
$\omega_1 = \alpha\omega_2$ and $\tan\phi = \omega_2/\omega_1\sqrt{1 + \alpha^{-2}}$.   
This integral converges if $x < 7/4$. 

With $x < 1$, we proceed as above and get that 
$I_b \sim \int\omega_{1}^{3/2 - x}d\omega_1d\phi.$
Convergence occurs when $x < 5/2$ that in agreement with the assumption $x < 1$. Therefore the integral is convergent 
for $x < 1$.

Now let us consider  case (c): $\omega_1 \approx \omega_2 \gg \omega_3 \sim \omega $. 
Then we have  $S = \sqrt{\omega}$ and
\begin{equation}\label{E6}
I_c \sim \int\omega_{1}^{- 2x}\left(\omega^x + \omega_1^x - \omega_3^x - (\omega + \omega_1 - \omega_3)^x\right)d\omega_1d\omega_3.
\end{equation} 
Again with Taylor expanding, we get
$I_c \sim \int\omega_{1}^{- 2x}\left(\omega^x - \omega_3^x - \omega_1^{x-1}x(\omega - \omega_3)\right)d\omega_1d\omega_3.$
For $x > 1$, the integral is convergent. 
For $x \leq 1$  the integral is convergent when $x > 1/2$.

Finally, consider  case (d): $\omega_1$, $\omega_2$, $\omega_3 \gg \omega $. Therefore $S = \sqrt{\omega}$ and we have
\begin{equation}\label{7a}
I_d \sim \int 
(\omega_{1}\omega_{2}\omega_{3})^{-x}
\left({\omega_1}^{x}  - \omega_{2}^{x} - \omega_{3}^{x}\right)\delta(\omega_1 - \omega_2
- \omega_3)d\omega_1d\omega_2d\omega_3.
\end{equation}
Taylor expanding the bracket $({\omega_1}^{x}  - \omega_{2}^{x} - \omega_{3}^{x}) = \omega_2^x[(1 + \alpha^{-1})^x - 1 - \alpha^{-x}]$
with $\omega_2 = \alpha\omega_3$ and $\omega_1 = \omega_2  + \omega_3 = \omega_2(1 + \alpha^{-1})$, we get
$I_d \sim \int\omega_{2}^{-2x + 1}d\omega_2.$
The integral is convergent iff $x > 1$. 

{ In summary, the integral is convergent when $1<x <3/2$ and divergent otherwise. Divergence for $x\ge 3/2$ occurs at  $\omega_1 \ll \omega_2$, $\omega_3 \sim \omega $, and for $x\le 1$ at $\omega_1$, $\omega_2$, $\omega_3 \gg \omega $. }

\section*{Appendix 2. Mappings of the subdomains of decomposition of $\Delta_{\eta}$ to the reference square and stripes}\label{A2}

Below the explicit formulas of the mappings $\mathfrak{F}^{-1}_k:R_{sq}\rightarrow$T$_k$ are given together with the Jacobians $J_{Tk}$
of these mappings that are also needed for computing the integrals. Here $R_{sq}=\{(y,z):-1\leq y,z\leq 1\}$ is a reference square,
T$_k$, $k=1,...,5$ denote the trapeze subdomains demonstrated in Fig.~\ref{fig2}, wherein $(\eta_2,\eta_3)\in$T$_k$.
\begin{itemize}
\item For T$_1$ the mapping $\mathfrak{F}^{-1}_1$ is determined as
\begin{equation*}
\eta_2=\frac{\eta_{min}}{2}(y+1),~~~\eta_3=\frac{1}{a(y)}\biggl(\frac{\eta}{2}z+\frac{2\eta_{max}+\eta}{2}\biggr)-\frac{\eta_{min}\eta_{max}(z+1)}{2\eta},~~~J_{T1}=\frac{\eta_{min}\eta}{4a(y)},
\end{equation*}
where $a(y)=\dfrac{2\eta}{2\eta-\eta_{min}(y+1)}$.
\item For T$_2$ the mapping $\mathfrak{F}^{-1}_2$ is determined as
\begin{equation*}
\eta_2=\frac{\eta_{min}}{2}(y+1),~~~\eta_3=\frac{1}{a(y)}\biggl(\frac{\eta_{max}-\eta}{2}z+\frac{\eta_{max}+\eta}{2}\biggr)-\frac{\eta_{min}\eta_{max}(y-1)}{2(\eta_{max}-\eta)},
\end{equation*}
$J_{T2}=\dfrac{\eta_{min}(\eta_{max}-\eta)}{4a(y)},$ where $a(y)=\dfrac{2(\eta_{max}-\eta)}{2(\eta_{max}-\eta_{min}-\eta)+\eta_{min}(y+1)}$.
\item For T$_3$ the mapping $\mathfrak{F}^{-1}_3$ is determined as
\begin{equation*}
\eta_2=\frac{\eta-\eta_{min}}{2a(z)}y+\frac{1}{4}\eta_{min}(z+1)+\frac{\eta}{2},~~~\eta_3=-\frac{\eta-\eta_{min}}{2a(z)}y+\frac{1}{4}\eta_{min}(z+1)+\frac{\eta}{2},
\end{equation*}
$J_{T3}=\dfrac{\eta_{min}(\eta-\eta_{min})}{4a(z)},$ where $a(z)=\dfrac{2(\eta-\eta_{min})}{2\eta_{min}(z-3)+2\eta}$.
\item For T$_4$ the mapping $\mathfrak{F}^{-1}_4$ is determined as
\begin{equation*}
\eta_2=\frac{1}{a(z)}\biggl(\frac{\eta_{max}-\eta}{2}y+\frac{\eta_{max}+\eta}{2}\biggr)-\frac{\eta_{min}\eta_{max}(z-1)}{2(\eta_{max}-\eta)},~~~\eta_3=\frac{\eta_{min}}{2}(z+1),
\end{equation*}
$J_{T4}=\dfrac{\eta_{min}(\eta_{max}-\eta)}{4a(z)},$ where $a(z)=\dfrac{2(\eta_{max}-\eta)}{2(\eta_{max}-\eta_{min}-\eta)+\eta_{min}(z+1)}$.
\item For T$_5$ the mapping $\mathfrak{F}^{-1}_5$ is determined as
\begin{equation*}
\eta_2=\frac{1}{a(z)}\biggl(\frac{\eta}{2}y+\frac{2\eta_{max}+\eta}{2}\biggr)-\frac{\eta_{min}\eta_{max}(z+1)}{2\eta},~~~\eta_3=\frac{\eta_{min}}{2}(z+1),~~~J_{T5}=\frac{\eta_{min}\eta}{4a(z)},
\end{equation*}
where $a(z)=\dfrac{2\eta}{2\eta-\eta_{min}(z+1)}$.
\end{itemize}
Below the explicit formulas of the mappings $\mathfrak{F}^{-1}:\tilde{R}_{st}\rightarrow\Omega$ together with the Jacobians $J_{\Omega}$
of these mappings are given. Here $\tilde{R}_{st}=\{(\tilde{y},\tilde{z}):
\tilde{y}\in[0,\infty), \tilde{z}\in[-1,1]\}$  is a reference
stripe, $\Omega$ is one of the unbounded subdomains $\psi_{1,2}$, $\varphi_1$ demonstrated in Fig.~\ref{fig2}, wherein
$(\eta_2,\eta_3)\in\Omega$.
\begin{itemize}
\item For $\psi_1$ the mapping $\mathfrak{F}^{-1}_{\Omega}$ is determined as
$$
\eta_2=\frac{2}{\eta_{min}}J_{\psi1}\tilde{y}+\eta_{max},~~~\eta_3=\frac{\eta_{min}}{2}(\tilde{z}+1),~~~J_{\psi1}=\frac{\eta_{min}(2\eta+2\eta_{max}-\eta_{min}(\tilde{z}+1))}{4\eta_{max}}.
$$
\item For $\psi_2$ the mapping $\mathfrak{F}^{-1}_{\Omega}$ is determined as
$$
\eta_2=\frac{2}{\eta-\eta_{min}}J_{\psi2}\tilde{y}+\eta_{max},~~~\eta_3=\frac{\eta-\eta_{min}}{2}\bigg(\tilde{z}+\frac{\eta+\eta_{min}}{\eta-\eta_{min}}\bigg),
$$
$J_{\psi2}=\dfrac{(\eta-\eta_{min})(2\eta+2\eta_{max}-\eta_{min}(\tilde{z}+1))}{4\eta_{max}}$.
\item For $\varphi_1$ the mapping $\mathfrak{F}^{-1}_{\Omega}$ is determined as
$$
\eta_2=\tilde{y}+\eta_{max},~~~\eta_3=\frac{\eta_{max}-\eta}{2}\bigg(\tilde{z}+\frac{\eta_{max}+\eta}{\eta_{max}-\eta}\bigg),~~~J_{\varphi1}=\frac{\eta_{max}-\eta}{2}.
$$
\end{itemize}
By analogy the formulas of the mappings of the reference stripe $\widehat{R}_{st}=\{(\widehat{y},\widehat{z}): \widehat{y}\in[-1,1], \widehat{z}\in[0,\infty)\}$ to the unbounded subdomains $\psi_{3,4}$, $\varphi_2$ can be written as
\begin{itemize}
\item For $\psi_3$:
$$
\eta_2=\frac{\eta-\eta_{min}}{2}\bigg(\widehat{y}+\frac{\eta+\eta_{min}}{\eta-\eta_{min}}\bigg),~~~\eta_3=\frac{2}{\eta-\eta_{min}}J_{\psi3}\widehat{z}+\eta_{max},
$$
$J_{\psi3}=\dfrac{(\eta-\eta_{min})(2\eta+2\eta_{max}-\eta_{min}(\widehat{y}+1))}{4\eta_{max}}.$
\item For $\psi_4$:
$$
\eta_2=\frac{\eta_{min}}{2}(\widehat{y}+1),~~~\eta_3=\frac{2}{\eta_{min}}J_{\psi4}\widehat{z}+\eta_{max},~~~J_{\psi4}=\frac{\eta_{min}(2\eta+2\eta_{max}-\eta_{min}(\widehat{y}+1))}{4\eta_{max}}.
$$
\item For $\varphi_2$:
$$
\eta_2=\frac{\eta_{max}-\eta}{2}\bigg(\widehat{y}+\frac{\eta_{max}+\eta}{\eta_{max}-\eta}\bigg),~~~\eta_3=\widehat{z}+\eta_{max},~~~J_{\varphi2}=\frac{\eta_{max}-\eta}{2}.
$$
\end{itemize}

\section*{Appendix 3. The derivation of integrals of the form~(\ref{sing1})  and  analysis of the error of quadrature formula for it}

We show how the following integral appears:
\begin{equation}
\label{sing1A}
\lambda(y)=\int_{-1}^1 (z + 1 + \varepsilon)^q{h}(y,z)dz,~~q \in \Bbb R, \quad  0 \leq \varepsilon\ll 1
\end{equation}
under the following two combinations of the parameters $q$ and $\varepsilon$: $q = 1/2$, $\varepsilon=0$ and $q \in \mathbb{R}$, $\varepsilon > 0$ is small.

Consider the integral operator $A(f,\eta)$. In the case of integration over the subdomain $\Delta_9$ (see Fig.~\ref{fig2}) we have 
\begin{equation}
\label{EA1}
A_{\Delta_9}(f,\eta)=\eta^{-1/2}\int\limits_{\Delta_9}\sqrt{\eta_3}C_0f(\eta_2)f(\eta_2+\eta_3-\eta)d\eta_2d\eta_3,
\end{equation}
where $C_0=f(\eta_{min})$ (see \eqref{C2_1}). Changing the variables in accordance with \eqref{C5} (see also the last column of Table~\ref{tabTriangPar}) for the domain $\Delta_9$
we have
\begin{equation}\label{EA2}
\begin{aligned}
&\eta_2=\frac{\eta_{min}}{8}(y+1)(3-z)+\eta,~~~\eta_3=\frac{\eta_{min}}{8}(z+1)(3-y),\\
&\eta_2+\eta_3-\eta=\frac{\eta_{min}}{4}(z-1)(1-y)+\eta_{min}.
\end{aligned}
\end{equation}
Then
\begin{equation}
\label{EA2_2}
A_{\Delta_9}(f,\eta)=\eta^{-1/2}\int_{-1}^1\int_{-1}^1g_{\Delta_9}(y,z)dydz,
\end{equation}
where
\begin{equation}
\label{EA3}
\begin{aligned}
&g_{\Delta_9}(y,z) = C_0\sqrt{\frac{\eta_{min}}{8}(3-y)(z+1)}f\biggl(\frac{\eta_{min}}{8}(3-z)(y+1)+\eta\biggr)\cdot\\
&\cdot f\biggl(\frac{\eta_{min}}{4}(z-1)(1-y)+\eta_{min}\biggr)\frac{\eta_{min}}{16}(2-y-z).
\end{aligned}
\end{equation}
Therefore, for computing the integral $A_{\Delta_9}(f,\eta)$ by using the Fubini theorem we have to consider  
\begin{equation}
\label{EA4}
\lambda(y)=\int_{-1}^1 \sqrt{z+1}h(y,z)dz,~~~y\in[-1,1],
\end{equation}
where
\begin{eqnarray}
&&h(y,z)=C_1(y)f\big(C_2(y)(3-z)+\eta\big)f\big(C_3(y)(z-1)+\eta_{min}\big)(2-y-z),\\
&&C_1(y)=C_0\dfrac{\eta_{min}}{16}\sqrt{\dfrac{\eta_{min}}{8}(3-y)},\\
&&C_2(y)=\dfrac{\eta_{min}}{8}(y+1), ~~~C_3(y)=\dfrac{\eta_{min}}{4}(1-y).
\end{eqnarray}
Therefore, we get exactly \eqref{sing1A} with $q=1/2$, $\varepsilon=0$. Notice that in \eqref{EA4} function $h$ is analytic if $f$ is analytic. Similarly we obtain the problem of integration of the form \eqref{EA4} for all subdomains of $\Delta_{\eta}$ located inside the dashed region in Fig.~\ref{fig2}.

Consider now the case of integration over the subdomain $\rho_2$:
\begin{equation}
\label{EA5}
A_{\rho_2}(f,\eta)=\eta^{-1/2}\int\limits_{\rho_2}\sqrt{\eta_3}f(\eta_2)f(\eta_3)f(\eta_2+\eta_3-\eta)d\eta_2d\eta_3.
\end{equation}
In accordance with \eqref{E10_1} we have $f(\eta)\sim \tilde{C}+C\eta^{-p}$ as $\eta\rightarrow 0$, where $\tilde{C}$ and $C$ are constants, $p > 0$. With this, there exists a bounded function $r(\eta)$ such that
\begin{equation}
\label{EA6} 
f(\eta)=(\tilde{C}+C\eta^{-p})r(\eta),
\end{equation} 
$r(\eta) \sim 1 $ as $\eta \to 0$. Assuming that $\eta_{min}$ is close to zero we use the representation 
 The change of variables  $\rho_2 \to R_{sq}$ leads to the following presentation \eqref{EA6} for expanding $f(\eta_3)$ in \eqref{EA5}
\begin{equation}
\label{EA7_0}  
A_{\rho_2}(f,\eta)=\eta^{-1/2}(I_1+I_2),
\end{equation} 
where
\begin{equation}
\label{A3_1}
I_1=\int_{-1}^1\int_{-1}^1\tilde{C}\frac{\eta-\eta_{min}}{2}\sqrt{z+1+\frac{2\eta_{min}}{\eta-\eta_{min}}}\xi(y,z)dydz,
\end{equation}
\begin{equation}
\label{A3_2}
I_2=\int_{-1}^1\int_{-1}^1C\biggl(\frac{\eta-\eta_{min}}{2}\biggr)^{p+1/2}\biggl(z+1+\frac{2\eta_{min}}{\eta-\eta_{min}}\biggr)^{p+1/2}r(\eta_3(z))\xi(y,z)dydz.
\end{equation}
Here
$$
\xi(y,z)=f\biggl(\frac{\eta_{max}-\eta}{2}y+\frac{\eta_{max}+\eta}{2}\biggr)f\biggl(\frac{\eta_{max}-\eta}{2}y+\frac{\eta-\eta_{min}}{2}z+\frac{\eta_{max}+\eta_{min}}{2}\biggr)C_{\xi},
$$
$$
\eta_3(z)=\frac{\eta-\eta_{min}}{2}(z+1)+\eta_{min},~~~C_{\xi}=\frac{(\eta-\eta_{min})(\eta_{max}-\eta)}{4}.
$$
Introducing in \eqref{A3_1}, \eqref{A3_2} the notation $\varepsilon=\dfrac{2\eta_{min}}{\eta-\eta_{min}}$, we can write 
\begin{equation}
\label{EA7}  
I_k=\int_{-1}^1\int_{-1}^1 (z+1+\varepsilon)^q\tilde{h}(y,z)dy dz,~~~k\in\{1,2\},
\end{equation}
where 
\begin{itemize}
\item $\tilde{h}(y,z)=\tilde{C}\dfrac{\eta-\eta_{min}}{2}\xi(y,z)$ and $q=1/2$ for $k=1$;
\item  $\tilde{h}(y,z)=C\biggl(\dfrac{\eta-\eta_{min}}{2}\biggr)^{p+1/2}\xi(y,z)r(\eta_3(z))$ and $q=p+1/2$ with a priori unknown $p$ for $k=2$. 
\end{itemize}
Hence, the integral $I_1$ corresponds to \eqref{EA4}, and the integral $I_2$ -- to the case of arbitrary $q$ and small positive $\varepsilon$ in \eqref{sing1A}.\\[3mm]

For the integral of the form \eqref{EA4} in section~\ref{feature3} the quadrature formula \eqref{C8} has been proposed. Let us now give the proof of the assertion~\ref{theo2} stating the exponential decay of the error of \eqref{C8}. We also provide a numerical verification of these results, see Fig.~\ref{fig4}.\\

{\it Proof of the proposition~\ref{theo2}}.~~~
Consider the best polynomial approximation $P_M^b(z)$ of $h_k(z)=h(y_k,z)$ such that
$$
\|h_k(z)-P_M^b(z)\|=\inf_{q(z)\in\mathcal{P}_{M}}\|h_k(z)-q(z)\|=E_{M},
$$
where $\mathcal{P}_{M}$ is the space of polynomials of the degree $\leq M$. With this and taking into account the formula \eqref{C3},
we have 
\begin{equation}
\label{SemFa}
\int\limits_{-1}^1(1\pm z)^{q}P_M^b(z)dz-\sum\limits_{m=0}^{M}\tilde{a}_{mk}
\mathcal{T}_m^q=0,~~~\tilde{a}_{mk}=\frac{c_m}{\pi}\int\limits_{-1}^1\frac{P_M^b(z)T_m(z)}{\sqrt{1-z^2}}dz.
\end{equation}
Then following the notations of \eqref{C8} we get 
\begin{equation}
\label{est1}
\begin{aligned}
|I_h(y_k)-I_h^{M}&(y_k)|\leq\int\limits_{-1}^1(1\pm z)^{q}|P_M^b(z)-h_k(z)|dz+\sum\limits_{m=0}^{M}|\tilde{a}_{mk} -
a_{mk}||\mathcal{T}_m^q|\leq\\
&\leq E_M\biggl(1+\frac{2}{\pi}\int\limits_{-1}^1\frac{dz}{\sqrt{1-z^2}}\biggr)\int\limits_{-1}^1(1\pm z)^{q}dz=3\frac{2^{1+q}}{1+q}E_M,
\end{aligned}
\end{equation}
wherein we used that $|T_m(z)|\leq 1$. In accordance with \cite{bernstein1912best} and \cite{trefethen2019approximation}~Chapter~8, Theorem~8.2, the estimate of error of the best approximation of $h_k(z)$ is
\begin{equation}
\label{est2}
E_M\leq\frac{2L\rho^{-M}}{\rho-1},~~~\rho>1,
\end{equation}
where $\rho>1$ is the sum of semi-axes of the Bernstein ellipse  $\mathfrak{E}_{\rho}$~\cite{bernstein1912best}. 
The quantity $\rho$
characterizes the distance from $[-1,1]$ to the closest singular point of an analytic continuation of $h(y_k,z)$ in the
complex plain.  
The constant
$L$ is upper bound of the modulus of this analytic continuation i.e. $|h(y_k,z)|\leq L$ for $z\in \mathfrak{E}_{\rho}$.
Combining \eqref{est1} with \eqref{est2} we obtain the error estimate  of approximation in the formula \eqref{C8}
which leads to the exponential convergence.\qed

Fig.~\ref{fig4} shows an examples of numerical calculations   for  test functions $h(y,z)=h_1(z)=\exp(10z)$, $q=0.5$ and $q=-0.9$ with convergence of the formulas \eqref{C8}, \eqref{SemF} presented in the logarithmic scale.

\begin{figure}[ht]
\centering
\includegraphics[scale=0.33]{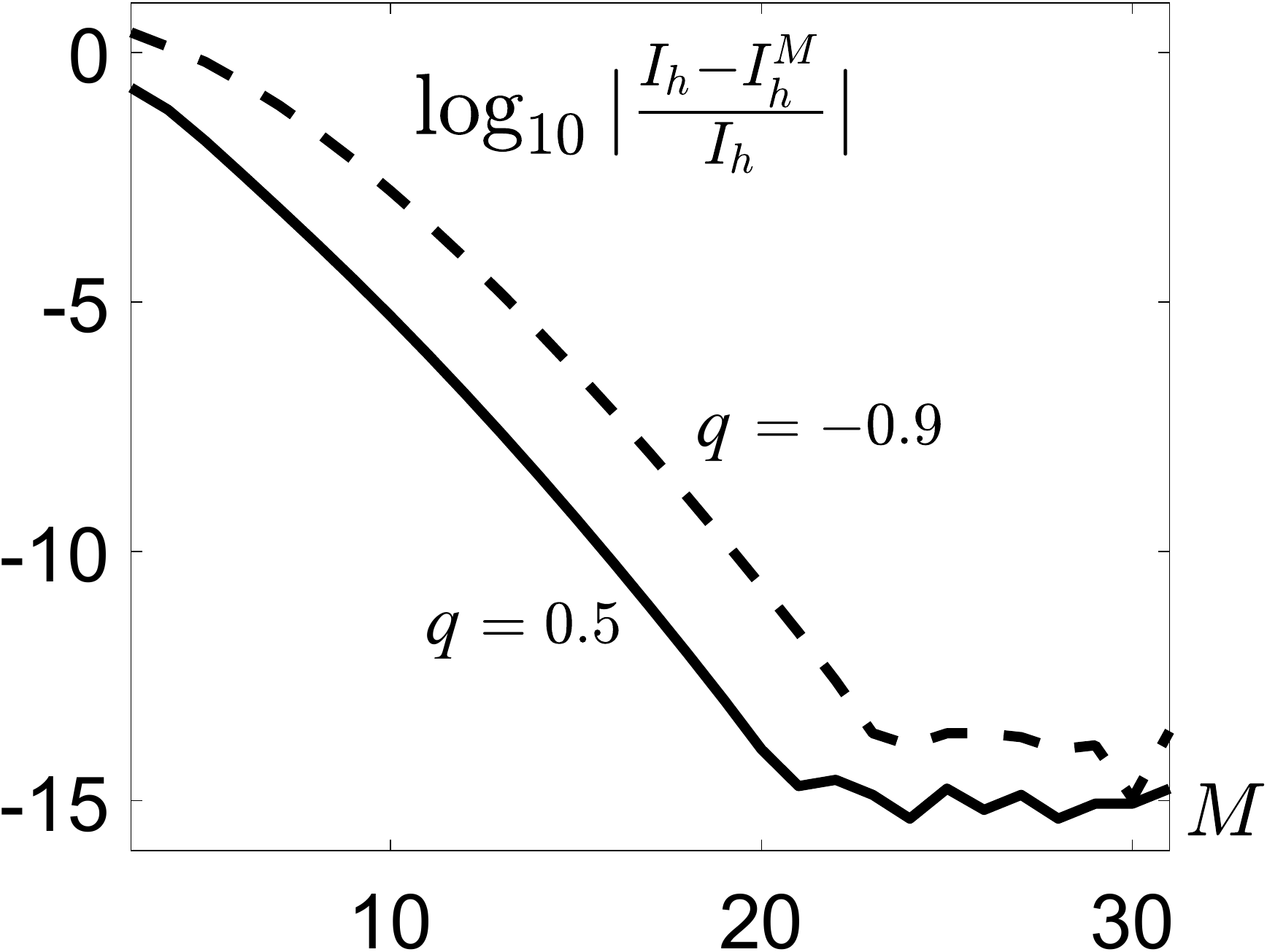}
\caption{Relative errors of integration of function $\sqrt{1+z}h_1(z)$ (solid line) and $(1+z)^{-0.9}h_1(z)$ (dashed line) in logarithmic scale, where $h_1(z)=\exp(10z)$}
\label{fig4}
\end{figure}

\section*{Appendix 4. Verification of the cubature formulas for computing the integrals $A(f,\eta)$ and $B(f,\eta)$}

To preform the calculations, we use the MATLAB environment. The following test function
$f_T(\eta)$ (see Fig.~\ref{fig5}), which is specified by the three parameters $\eta_{min}$, $\eta_{max}$ and $x$, is used:
\begin{equation}
\label{TestF1}
f_T(\eta)=\left\{
\begin{aligned}
&C_0,~0\leq\eta<\eta_{min},\\
&-\frac{a\eta^2+b\eta+1}{\eta^{x}},~\eta_{min}\leq\eta\leq\eta_{max},\\
&C_{\infty}\eta^{-x},~\eta>\eta_{max},
\end{aligned}
\right.
\end{equation}
where

$
a=\dfrac{x}{(2-x)\eta_{min}^2-2(1-x)\eta_{max}\eta_{min}},
~b=-2a\eta_{max},$

$C_0=-\dfrac{a\eta_{min}^2+b\eta_{min}+1}{\eta_{min}^{x}},~C_{\infty}=-(a\eta_{max}^2+b\eta_{max}+1).
$

The test function $f_T(\eta)$ imitates important properties of solutions of the problem \eqref{C1}, \eqref{C2}.
Specifically, $f_T(\eta)$ has the shelf-type behavior for small $\eta$ and the power law $\eta^{-x}$ for 
$\eta > \eta_{max}$, the function growth as $\eta^{-x}$ for $\eta \approx \eta_{min}$  with $\eta_{min} \to 0$ that leads to the significant complication of computing the
corresponding integral for $\eta_{min} \ll \eta_{max}$.
\begin{figure}[ht]
\centering
\includegraphics[scale=0.37]{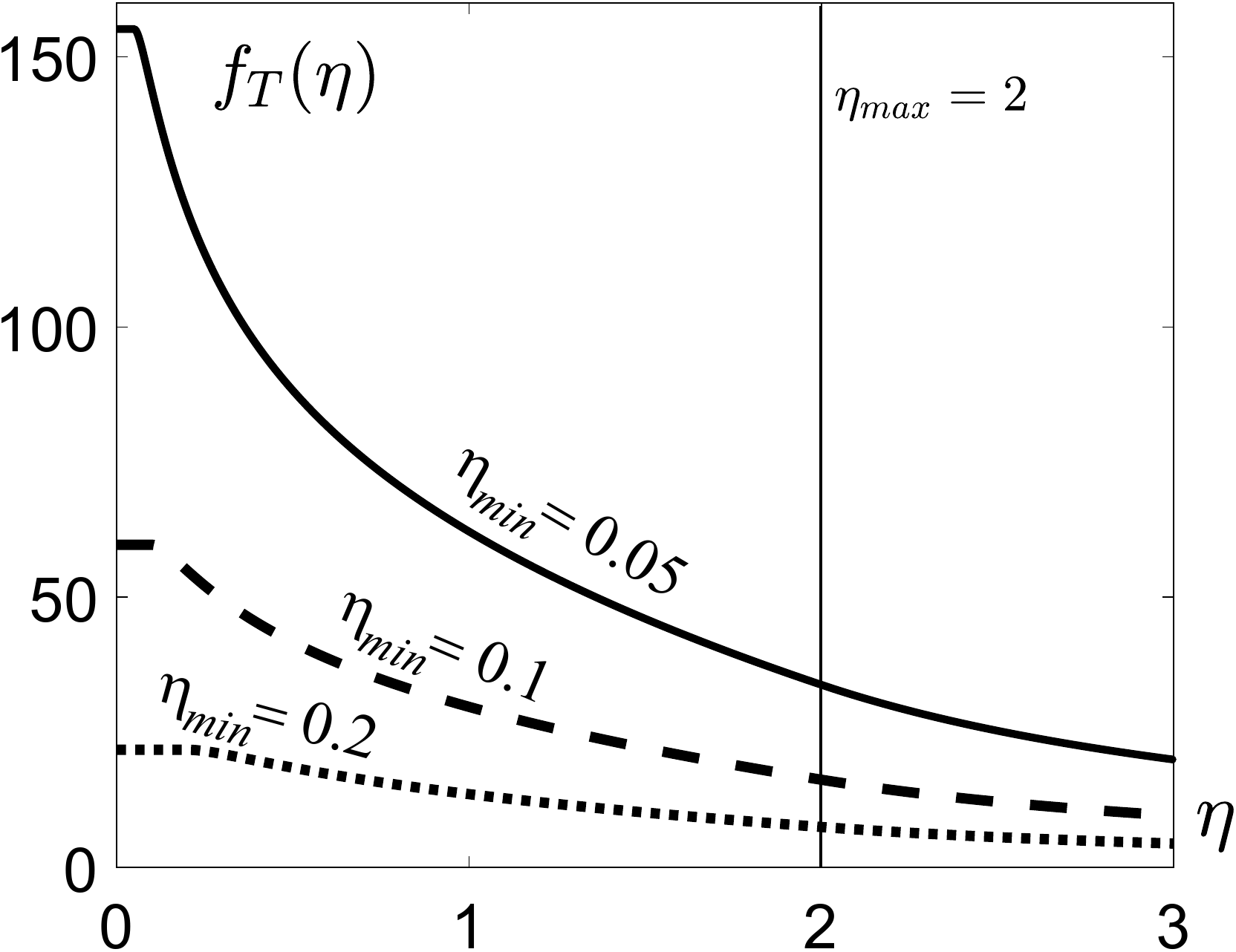}
\caption{Graphs of the test functions $f_T(\eta)$ with $x=1.3$, $\eta_{max}=2$ and different values of $\eta_{min}$: $\eta_{min}=0.05$
(solid line), $\eta_{min}=0.1$ (dashed line), $\eta_{min}=0.2$ (points) }
\label{fig5}
\end{figure}
We take $\eta=1.01$, $\eta_{min}=0.1$, $\eta_{max}=2$, $x=1.23$. Denote by $R_{ex}=A(f_T,\eta)+f_T B(f_T, \eta)$ the
exact value of the RHS of \eqref{C1} with $f=f_T$ and set $R_{ap}(N_t,M,N)$ to be the approximate value of the RHS obtained by 
computing.  The test function $f_T(\eta)$ is also approximated by the interpolation polynomial with $N$ Chebyshev nodes.  The symbol $N_t$  in $R_{ap}(N_t,M,N)$ denotes the number of nodes in the Double Exponential Formula \eqref{Tak1}, $M$ is the length of series
of the form \eqref{C3a} used for subdomains of the decomposition of $\Delta_{\eta}$. With this, the results of calculations presented in Fig.~\ref{fig6} show the dependence of
the numerical error on $N_t$, $M$ and $N$ for the specified values of $\eta$, $\eta_{min}$,
$\eta_{max}$, $x$. Here, instead of $R_{ex}$ we take the corresponding approximation obtained with $N_t=801$,
$M=81$, $N=71$.
\begin{figure}[ht]
\centering
\includegraphics[scale=0.42]{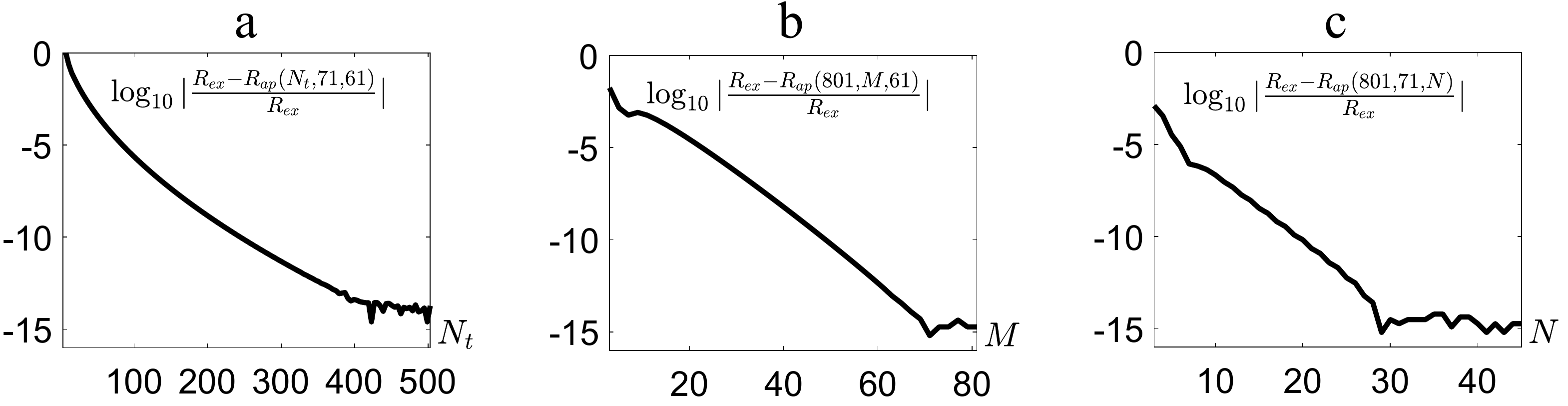}
\caption{Dependence of the relative error of 
 values of RHS on the value of $N_t$ (a), $M$ (b), $N$(c)}
\label{fig6}
\end{figure}
The exponential convergence is clearly observed in the panels b and c, but the convergence presented in panel a is sub-exponential. This is caused by presence of the factor $\frac{1}{\log N_t}$ in estimate \eqref{C7}. However the number of operations of the presented cubature formulas has linear dependence on $N_t$. Then taking for example $N_t>200$ gives fast and very accurate integration.

The next step of verification was the comparison with results obtained in Wolfram Mathematica (WM).
Firstly we compute $A(f_T,\eta)+f_T B(f_T, \eta)$ by WM. Then we repeat these calculations by  the  proposed method (PM) for different values of the parameters, $\eta_{min}$, $\eta$, $x$ and  $\eta_{max}=2$. Table~\ref{tab0} shows the comparison of the results obtained by the WM and PM. The time consumption is also presented. The results obtained by the PM have been checked with doubling the values of $N_t$, $M$ and $N$. The relative deviation of the results obtained for current and doubled values of $N_t$, $M$ and $N$ is always less then $2.27\times 10^{-11}$.

\begin{table*}[!hbt]
\begin{small}
\caption{Results of integration by the method proposed  (PM) and
by Wolfram Mathematica (WM)} \label{tab0}
\begin{center}
\resizebox{\textwidth}{!}{%
   \begin{tabular}{c|c|c|c|c|c|c|c}  \hline
$\eta_{min}$   & $x$ & $\eta$ & PM result & WM result & $|$PM-WM$|$/PM & PM time (s) & WM time (s)\\[1ex]
\hline 0.5       & 1.5      & 0.501       & 39.9957739778432 &
39,9957739480505 & $1.45\times 10^{-10}$ & 0.36        & 37.2\\
0.5     &   1.23 & 0.7 & 187.95695591569  & 187.956955768542 &
$7.83\times 10^{-10}$  &
0.42 & 40.9\\
0.5 & 1.23   & 1.49 & 41.2463746906135 & 41.2463747732553 &
$2\times 10^{-9}$  & 0.42 &  43.9\\
0.1 &   1.5   &  0.8   &  4842.07813448566    &   4842.07816293161 & $5.87\times 10^{-9}$  & 0.62    & 43.3\\
0.1  & 1.23  &   1.01 & 46536.0787163029 & 46536.0787198942 &
$7.71\times 10^{-11}$ & 0.52 & 45.2
\\
0.1  & 1.03  &   2 & 3473744.4765 & 3473276.9800 &
$1.35\times 10^{-4}$ & 0.067 & 48.3
\\
0.01 & 1.16 & 1.501 & 94784029.8204267 & 94783980.7188906 &
$5.18\times 10^{-7}$ & 0.22 & 46.8\\
0.01 & 1.35 & 0.99 & 12874999.2377083 & 12874982.1165897 &
$1.33\times 10^{-6}$ & 0.46 & 46.6 \\
0.01 & 1.5 & 1.99 & -2427937.8611 & -2428064.0375 &
$5.2\times 10^{-5}$ & 0.49 & 45.6\\[1ex] \hline
\end{tabular}}
\end{center}
\end{small}
\end{table*}
From the data in Table~\ref{tab0} we can conclude that:  1) the
results obtained by the PM and WM are rather close that ensures the correctness of the designed methods; 2) as shows the analysis of convergence given in Fig.~\ref{fig6} and doubling of number of nodes, the results obtained by PM are always more accurate than those obtained by WM; 3) the time performance by WM is rather poor even in the simplest case of large $\eta_{min}$; 4) in the case when $\eta_{min}$ is decreasing, the accuracy of WM also decreases, however for the values $\eta_{min}=0.01$ it remains rather high.

\section*{Appendix~5. Approximation of $f(\eta)$ and its derivatives}

To approximate $f(\eta)$,  we first use the linear change of variable $\mathfrak{L}:~[\eta_{min},\eta_{max}]\rightarrow [-1,1]$ such that $\mathfrak{L}(\eta_{min})=-1$, $\mathfrak{L}(\eta_{max})=1$ with $y=\mathfrak{L}\eta$, $y\in[-1,1]$. The function $\tilde{f}(y)=f(\mathfrak{L}^{-1}y)$ is defined on $[-1,1]$ and the interpolation polynomial with the Chebyshev nodes $y_1,...,y_N\in[-1,1]$, $y_j=\cos\frac{(2j-1)\pi}{2N}$,
\begin{equation}
\label{Apr1}
\begin{split}
p(\tilde f,y)=\sum\limits_{j=1}^N\frac{w(y,y_j)T_N(y)}{(y-y_j)T_N'(y_j)}\tilde f(y_j)&+w_{-1}(y)\tilde f(-1)+w_{1}(y)\tilde f(1)+\\
&+v_{-1}(y)\tilde f'(-1)+v_{1}(y)\tilde f'(1),
\end{split}
\end{equation}
will be used to approximate $\tilde{f}(y)$. We will omit the tilde symbol in the notation $\tilde{f}(y)$. Here  
$T_N(y)$ is the Chebyshev polynomial of $N$-degree and the functions $w$, $w_{\pm 1}$ and $v_{1}$ are as follows:
$$
w(y,y_j)=\frac{(1-y^2)^2}{(1-y_j^2)^2},
$$
$$
w_{\pm 1}(y)=0.25(\pm 1)^N(1\pm y)^2\{1+(1\mp y)(1+ N^2)\}T_N(y),~~~
v_1(y)=-0.25(1+y)(1-y^2)T_N(y).
$$
In view of the boundary  conditions \eqref{C2}, we have 
\begin{equation}\label{Apr1c}
f'(-1)  =  0, ~~~
f'(1)  =  \frac{d\mathfrak{L}^{-1}}{dy}\frac{df}{d\eta}\bigg|_{\eta  \equiv  \eta_{max}}=-d\frac{x}{\eta_{max}}f(1),
\end{equation}
where 
\begin{equation}\label{Apr1e}
d=\frac{d\mathfrak{L}^{-1}}{dy}.
\end{equation}
With this, instead of $w_{1}(y)f(1)+v_{1}(y)f'(1)$ in \eqref{Apr1} we will write $\hat{w}_1(y)f(1)$, where
\begin{equation}\label{Apr1f}
\hat{w}_1(y)=w_1(y)-d\frac{x}{\eta_{max}}v_1(y).
\end{equation}
Further, we use the notations  
\begin{equation}\label{Apr1g}
y_0=-1,~y_{N+1}=1,~f_j=f(y_j),~~~s=\sqrt{1-y^2},~s_j=\sqrt{1-y_j^2},~~~j=0,...,N+1.
\end{equation}
To approximate the derivatives of $f'(y)$ and $f''(y)$, we differentiate the formula \eqref{Apr1}:
\begin{equation}
\label{P1}
\begin{split}
p'(f,y)=\sum\limits_{j=1}^{N}\frac{(-1)^{j-1}s^2}{Ns_j^3}&\left\{\frac{4yy_j-3y^2-1}{(y-y_j)^2}T_N(y)+\frac{s^2}{y-y_j}T_N'(y)\right\}f_j+\\
&+\hat{w}_1'(y)f_{N+1}+w_{-1}'(y)f_0,
\end{split}
\end{equation}
\begin{equation}
\label{P2}
\begin{split}
p''(f,y)=&\sum\limits_{j=1}^{N}\frac{(-1)^{j-1}}{Ns_j^3}\left\{\biggl(\frac{2(1-y^4)-4y_j^2s^2}{(y-y_j)^3}+\frac{8y^2-N^2s^2}{(y-y_j)}\biggr)T_N(y)-\right.\\
&~~~~~~~~~~~~~\left.-s^2\frac{2s^2+7y(y-y_j)}{(y-y_j)^2}T_N'(y)\right\}f_j+\hat{w}_1''(y)f_{N+1}+w_{-1}''(y)f_0.
\end{split}
\end{equation}
As a result, we get 
\begin{equation}
\label{PD}
f'(y_i)\approx p'(f,y_i)=\sum\limits_{j=0,j\neq
i}^{N+1}a_{ij}f_j-\nu_if_i,~~~f''(y_i)\approx p''(f,y_i)=\sum\limits_{j=0,j\neq
i}^{N+1}b_{ij}f_j-\mu_if_i,
\end{equation}
where
$a_{ij}=(-1)^{i+j}\dfrac{s_i^3}{s_j^3(y_i-y_j)}$ for
$i\neq j$,~~~~~~$i,j=1,...,N$,\\

$a_{i0}=\dfrac{(-1)^{i-1}N(y_i+1)^2}{4s_i}(N^2-y_i(N^2+1)+2)+(-1)^{i}N(y_i+1)s_id\dfrac{x}{4\eta_{max}}$,~~~ $i=1,...,N$,\\

$a_{i~N+1}=\dfrac{(-1)^{i+N-1}N(y_i-1)^2}{4s_i}(N^2+y_i(N^2+1)+2)$,
~~~ $i=1,...,N$,\\

$a_{0j}=a_{N+1~j}=0$, $j=0,...,N+1$,~~
$\nu_{i}=-\dfrac{7y_i}{2s_i^2}$, $i=1,...,N$,~~
$\nu_{0}=d\dfrac{x}{\eta_{max}},~~\nu_{N+1}=0$;\\[3mm]

$b_{ij}=(-1)^{i+j-1}\dfrac{2s_i^3+7y_is_i(y_i-y_j)}{s_j^3(y_i-y_j)^2}$
as
$i\neq j$,~~~~~~$i,j=1,...,N$,\\

$b_{i0}=c_1\biggl([-N^2+5y_i(N^2+1)-4]y_i^2+[N^2-y_i(N^2+1)+2]y_i+2N^2-6y_i(N^2+1)+6+(5y_i-2)s_i^2 d\dfrac{x}{\eta_{max}}\biggr)$,\\

$b_{i~N+1}=c_2\biggl(-[N^2+5y_i(N^2+1)+4]y_i^2-[N^2+y_i(N^2+1)+2]y_i+2N^2+6y_i(N^2+1)+6\biggr)$,\\

where $c_1=\dfrac{(-1)^{i-1}N(y_i+1)}{4s_i^3}$,~~~
$c_2=\dfrac{(-1)^{i+N-1}N(y_i-1)}{4s_i^3}$,~~~ $i=1,...,N$;\\

$b_{0j}=\dfrac{8(-1)^{j-1}}{Ns_j^3(1-y_j)}$,~~~
$b_{N+1~j}=\dfrac{8(-1)^{j+N}}{Ns_j^3(1+y_j)}$, $j=1,...,N$, \\

$b_{0\ N+1}=\dfrac{(-1)^N(2N^2+3)}{2}$,~~~$b_{N+1\ 0}=b_{0\ N+1}-(-1)^Nd\dfrac{x}{\eta_{max}}$,\\

$\mu_i=-\dfrac{N^2+26}{3s_i^2}+\dfrac{5}{s_i^4}$, $i=1,...,N$,
~~~ $\mu_{N+1}=-\dfrac{N^2(5N^2+7)}{3}-\dfrac{3}{2}$,

$\mu_{0}=\mu_{N+1}+2(N^2+1)d\dfrac{x}{\eta_{max}}$.\\

Denote $e_j=\mathfrak{L}^{-1}(y_j)$, $j=0,...,\mathcal{N}$, $\mathcal{N}=N+1$ and then $f_j=\tilde{f}(y_j)=f(e_j)$. 
Consider the following vectors

\noindent$\mathbf{F}=\left(\begin{array}{c} f_0\\
f_1\\
\vdots\\
f_{\mathcal{N}} \end{array}\right),\ \mathbf{F}_{\eta}=\left(\begin{array}{c} (f_{\eta})_0\\
(f_{\eta})_1\\
\vdots\\
(f_{\eta})_{\mathcal{N}} \end{array}\right)=\left(\begin{array}{c} f_{\eta}({e}_0)\\
f_{\eta}(e_1)\\
\vdots\\
f_{\eta}(e_{\mathcal{N}}) \end{array}\right),\ \mathbf{F}_{\eta\eta}=\left(\begin{array}{c} (f_{\eta\eta})_0\\
(f_{\eta\eta})_1\\
\vdots\\
(f_{\eta\eta})_{\mathcal{N}} \end{array}\right)=\left(\begin{array}{c} f_{\eta\eta}(e_0)\\
f_{\eta\eta}(e_1)\\
\vdots\\
f_{\eta\eta}(e_{\mathcal{N}}) \end{array}\right)$\\

\noindent and $(N+2)\times (N+2)$ matrices

\noindent$\mathcal{A}=\mathcal{A}(x)=\dfrac{1}{d}\begin{pmatrix}
\nu_0 & a_{01}&...& a_{0~\mathcal{N}}\\
 a_{10}& \nu_1&...& a_{1~\mathcal{N}}\\
 \vdots &  \vdots &  ...&  \vdots\\
  a_{\mathcal{N}~0}& a_{\mathcal{N}~1}&...&\nu_{\mathcal{N}}
 \end{pmatrix}$,\ \ \
$ \mathcal{B}=\mathcal{B}(x)=\dfrac{1}{d^2}\begin{pmatrix}
  \mu_0& b_{01}&...& b_{0~\mathcal{N}}\\
 b_{10}& \mu_1&...& b_{1~\mathcal{N}}\\
 \vdots &  \vdots &  ...&  \vdots\\
  b_{\mathcal{N}~0}& b_{\mathcal{N}~1}&...& \mu_{\mathcal{N}}
 \end{pmatrix}$.\\
For $\mathbf{F}_{\eta}$ and $\mathbf{F}_{\eta\eta}$ the following approximate formulas hold: 
\begin{equation}
\label{AprMatr0} \mathbf{F}_{\eta}\approx\mathcal A\mathbf{F},\ \ \ \mathbf{F}_{\eta\eta}\approx\mathcal B\mathbf{F}.
\end{equation}
Notice that the elements $a_{i0}$, $i=1,...,N$, $\nu_0$ of the matrix $\mathcal{A}$  and the elements $b_{i0}$, $i=1,...,N+1$, $\mu_0$
of the matrix $\mathcal{B}$  depend on the spectral parameter $x$. These elements have to be recomputed on each iteration of the algorithms. 

Approximation~\eqref{AprMatr0} goes back to \cite{gottlieb1984theory}, where the similar formulas were obtained for much simpler interpolations. In some sources these formulas are called ``differentiation matrix technique''. Essential novelty of the formulas derived above consists in that they automatically satisfy the boundary conditions of the problem under consideration. The later is fundamentally important for posing and solving the spectral problems as in section~\ref{S4_2}. Such kind of approximations accounting for boundary conditions have been developed in \cite{semisalov2014non} on the basis of ideas from \cite{babenko2002fundamentals}.
Error estimates for \eqref{Apr1} and \eqref{PD} can be obtained by using the Cauchy interpolation theorem, see \cite{babenko2002fundamentals}, chapter~3,~section~3,~item~6.

\section*{Funding}
Sergey Nazarenko was supported by the Chaire  D'Excellence IDEX (Initiative of Excellence) awarded by
Universit\'e de la C\^ote d'Azur, France; the  European  Unions  Horizon  2020 research and innovation programme  in the framework of Marie Sklodowska-Curie HALT project (grant agreement No 823937); and the FET Flagships PhoQuS project (grant agreement No 820392).
Sergey Nazarenko and Boris Semisalov were supported by the
Simons Foundation Collaboration grant Wave Turbulence (Award ID 651471). Vladimir Grebenev and Boris Semisalov were partially supported by the ``chercheurs invit\'es" awards of the F\'ed\'eration Doeblin FR 2800, Universit\'e de la C\^ote d'Azur, France. Sergey Medvedev was partially supported by CNRS
``International Visiting Researcher" award and by the state assignment for fundamental research (FSUS-2020-0034).

\bibliography{references}

\begin{thebibliography}{99}
\bibitem{dyachenko1992optical} {\em A. Dyachenko, A.C. Newell, A. Pushkarev and V.E. Zakharov}
  Optical Turbulence: Weak Turbulence, Condensates and Collapsing Filaments in the Nonlinear Schrodinger Equation
{ Physica D}, 1992, {Vol. 57}, P. 96--160.


\bibitem{nazarenko2011wave} {\em S.~V. Nazarenko} Wave Turbulence. Springer-Verlag, Berlin Heidelberg, 2011.



\bibitem{semikoz1995kinetics} {\em D.~V.~Semikoz and I.~I.~Tkachev} Kinetics of Bose Condensation. 
Phys. Rev. Lett. 1995, Vol. 74(16). P.~3093--3097.

\bibitem{semikoz1997condensation} {\em D.~V.~Semikoz and I.~I.~Tkachev} Condensation of bosons in kinetic regime.  Phys. Rev. D, 1997, Vol. 55, 489--502.


\bibitemlacaze2001dynamical} {\em R.~Lacaze, P.~Lallemand, Y.~Pomeau, S.~Rica} Dynamical formation of a Bose--Einstein condensate. Physica D.
    Nonlinear Phenomena. 2001. Vol. 152. P. 779--786.

 \bibitem{connaughton2004kinetic} {\em C.~Connaughton, Y.~Pomeau} Kinetic theory and Bose--Einstein condensation. Comptes Rendus Physique. 2004.
Vol.~5, Is.~1. P. 91--106.

    
\bibitem{zel2002physics}
 {\em Zeldovich Ya B and Raizer Yu P} 2002 Physics   of   Shock-Waves   and   High-Temperature Hydrodynamics Phenomena (New York: Dover) p 944

\bibitem{galtier2000weak}
{\em
S. Galtier, S. V. Nazarenko, A. C. Newell, A. Pouquet},
A weak turbulence theory for incompressible magnetohydrodynamics,
	Journal of Plasma Physics, Vol. 63,
Issue number
	5,
Pages 
	447-488,
	Published - Jun 2000
 

\bibitem{connaughton2010dynamical} {\em C. Connaughton,  A. C. Newell}
Dynamical scaling and the finite-capacity anomaly in three-wave turbulence, Phys. Rev. E 81, 036303, Published 2 March 2010

\bibitem{connaughton2004warm}
{\em C. Connaughton, S. Nazarenko},
Warm cascades and anomalous scaling in a diffusion model of turbulence, 
 Physical review letters 92 (4), 044501, 2004

\bibitem{connaughton2003non}
{\em 
C. Connaughton, A. C. Newell, Y. Pomeau,} 
Non-stationary spectra of local wave turbulence,
Physica D: Nonlinear Phenomena Volume 184, Pages 64-85, 

\bibitem{grebenev2013self}
{\em
V. N. Grebenev, S. V. Nazarenko, S. B. Medvedev, I. V. Schwab, Y. A. Chirkunov,}
Self-similar solution in the Leith model of turbulence: anomalous power law and asymptotic analysis,
Journal of Physics A: Mathematical and Theoretical 47 (2), 025501

\bibitem{thalabard2015anomalous}
{\em
S. Thalabard, S. V. Nazarenko, S. Galtier, S. Medvedev,}
Anomalous spectral laws in differential models of turbulence,
Journal of Physics A: Mathematical and Theoretical 48 (28), 285501, 2015

\bibitem{galtier2019nonlinear}
{\em
 S. Galtier, S. V. Nazarenko, E. Buchlin, S. Thalabard,}
  Nonlinear diffusion models for gravitational wave turbulence,
Physica D: Nonlinear Phenomena 390, 84-88, 2019

\bibitem{bell2017self}
{\em 
N. K. Bell, V. N. Grebenev, S. B/ Medvedev, S. V. Nazarenko,}
Self-similar evolution of Alfven wave turbulence, 
Journal of Physics A: Mathematical and Theoretical 50 (43), 435501, 2017.

\bibitem{korotkevich2008numerical}
{\em A.O.Korotkevich, A.Pushkarev, D.Resio, V.E.Zakharov} Numerical verification of the weak turbulent model for swell evolution. European Journal of Mechanics - B/Fluids. Volume 27, Issue 4, 2008, Pages 361--387.

\bibitem{badulin2005self}
{\em S. I. Badulin, A. N. Pushkarev, D. Resio, V. E. Zakharov.} Self-similarity of wind-driven seas. Nonlin. Processes Geophys., 12, 891--945, 2005.
 
\bibitem{badulin2017ocean}
{\em S. I. Badulin, V. E. Zakharov.} Ocean swell within the kinetic equation for water waves. Nonlin. Processes Geophys., 24, 237--253, 2017.

\bibitem{polnikov1997nonlinear}
{\em V. G. Polnikov.}  Nonlinear energy transfer through the spectrum of gravity waves for the finite depth case. J. Phys. Oceanogr. 27, 1481-- 1491, 1997.

\bibitem{janssen2003nonlinear}
{\em  P.A.E.M. Janssen.}  Nonlinear four-wave interactions and freak waves. Journal of Physical Oceanography 33, 863--884, 2003.

\bibitem{resio1991numerical}
{\em Resio, D.T., Perrie, W.} 1991. A numerical study of nonlinear energy fluxes due to wave--wave interactions. Part 1: Methodology and basic results. Journal of Fluid Mechanics 223, 609--629.

\bibitem{van2006wrt}
{\em van Vledder, Gerbrant Ph.} The WRT method for the computation of non-linear four-wave interactions in discrete spectral wave models." Coastal Engineering 53.2--3 (2006): 223--242.

\bibitem{polnikov2002problem}
{\em Polnikov, V.G., Farina, L.} 2002. On the problem of optimal approximation for the four-wave kinetic integral. Nonlinear Processes in Geophysics 9, 497--512.

     \bibitem{hossain2014generalized} {\em M. Alamgir Hossain and Md. Shafiqul Islam} Generalized Composite Numerical Integration
Rule Over a Polygon Using Gaussian Quadrature. Dhaka Univ. J. Sci.
2014. 62(1): 25--29.

     \bibitem{takahasi1974double} {\em Hidetosi Takahasi and Masatake Mori} Double Exponential Formulas for
Numerical Integration. PM. RIMS, Kyoto Univ. 1974. 9: 721--741.

     \bibitem{trefethen2019approximation} {\em L. N. Trefethen} Approximation theory and approximation practice. SIAM, 2013.

     \bibitem{clenshaw1960method} {\em C. W. Clenshaw and A. R. Curtis} A method for numerical integration on an
automatic computer. Numer. Math. 1960. Vol.~2. P. 197--205.

	\bibitem{gentleman1972implementing} {\em W. M. Gentleman} Implementing Clenshaw--Curtis quadrature, II
computing the cosine transformation. Comm. ACM. 1972. Vol.~15. P.~343--346.

	\bibitem{weideman2007kink} {\em J. A. C. Weideman and L. N. Trefethen} The kink phenomenon in Fejer and
Clenshaw--Curtis quadrature. Numer. Math. 2007. Vol.~107. P~707--727.

\bibitem{bernstein1912best} {\em S.~N.~Bernstein} On the Best Approximation of Continuous Functions by Polynomials
of a Given Degree. Soobshch. Khar'kov Mat. Obshch. 1912, No. 13, pp. 49--144. (in Russian)

\bibitem{ralston2001first}  {\em A.~Ralston, Ph.~Rabinowitz} A First Course in Numerical Analysis (2nd ed.). New York: Dover Publications, 2001.

     \bibitem{tee2006rational} {\em T. W. Tee and L. N. Trefethen} A rational spectral collocation method with
adaptively transformed Chebyshev grid points. SIAM J. Sci. Comp.
2006. 28(5): 1798--1811.

\bibitem{gottlieb1984theory} {\em D. Gottlieb, M. Y. Hussaini and S. A. Orszag} Introduction: theory and
applications of spectral methods, in R. G. Voigt, D. Gottlieb and M. Y. Hussaini,
Spectral Methods for Partial Differential Equations, SIAM, 1984.

\bibitem{semisalov2014non} {\em B. V. Semisalov} Non-local algorithm of finding solution to the Poisson equation and its applications.
Zh. Vychisl. Mat. Mat. Fiz., 2014, Vol. 54, No. 7, p. 1110--1135 (in Russian)

\bibitem{blokhin2009numerical} {\em A.M. Blokhin,~A.S. Ibragimova} Numerical Method for 2D Simulation of a Silicon MESFET with a Hydrodynamical Model Based
    on the Maximum Entropy Principle. SIAM Journal on Scientific Computing. 2009. Vol.~31(3). P.~2015--2046.

   \bibitem{babenko2002fundamentals} {\em Babenko K.~I.} Fundamentals of numerical analysis. M. Glav. Red. Fiz.-Mat Lit. 1986. (in Russian)

    \bibitem{Salzer1972lagrangian} {\em H.~E.~Salzer} Lagrangian interpolation at the Chebyshev points $x_{n,\nu}=\cos(\nu \pi/n)$, $\nu=O(1)n$; sum unnoted advantages. The Computer Journal. 1972. Vol.~15, Is.~2. P. 156--159.


\bibitem{escobedo2014blow}{\em
    M. Escobedo and J. J. L. Velázquez } On the Blow Up and Condensation of Supercritical Solutions of the Nordheim Equation for Bosons. Communications in Mathematical Physics, volume 330, pages 331–365 (2014)
    
    \bibitem{galtier2017turbulence}{\em
   S. Galtier,  S.V. Nazarenko.} Turbulence of Weak Gravitational Waves in the Early Universe. Physical Review Letters. March 2017. 119(22).
    
    \bibitem{skipp2020wave}{\em J Skipp, V L'vov, S Nazarenko.} Wave turbulence in self-gravitating Bose gases and nonlocal nonlinear optics, 2020,
Physical Review A 102 (4), 043318


















\end{thebibliography}

\end{document}


\end{document}